\documentclass[11pt,onecolumn]{IEEEtran}
\usepackage{amsmath}
\usepackage{amssymb}
\usepackage[dvipdfmx]{graphicx}
\usepackage{ascmac}
\usepackage{multirow}
\usepackage[dvipdfmx]{color}

\ifCLASSINFOpdf
\else
\fi
\begin{document}

\sloppy

\title{Variable-Length Resolvability \\ for General Sources and Channels}

\author{Hideki~Yagi~~~~~Te~Sun~Han
\thanks{%
This research is supported by JSPS KAKENHI Grant Numbers JP16K06340 and 	26289119.}%
\thanks{H.\ Yagi is with the Dept.\ of Computer and Network Engineering, University of Electro-Communications, Tokyo, Japan (email: h.yagi@uec.ac.jp).}
\thanks{T.\ S.\ Han is with the National Institute of Information and Communications Technology (NICT), Tokyo, Japan (email: han@is.uec.ac.jp).}
}
\def\bf#1{\boldsymbol{#1}}
\def\vect#1{\boldsymbol{#1}}
\def\QED{\hfill$\Box$}
\definecolor{Bgreen}{rgb}{ .0, .55, .0 }
\definecolor{Red}{rgb}{ 1.0, .0, .0 }
\definecolor{Navy}{rgb}{ 0.0, .0, 1.0 }

\newcommand{\SrAlphabet}{{\mathcal{U}}}
\newcommand{\CdAlphabet}{{\mathcal{X}}}
\newcommand{\CoinSr}{{U^n}}
\newcommand{\TargetSr}{{X^n}}
\newcommand{\ApproxSr}{{\tilde{X}^n}}
\newcommand{\CoinSeq}{{\vect{u}}}
\newcommand{\TargetSeq}{{\vect{x}}}
\newcommand{\Tr}{\textrm{Tr}}
\newcommand{\E}{\mathbb{E}}	
\newcommand{\V}{\mathsf{V}}	
\newcommand{\ep}{\textrm{ep}}
\newcommand{\cl}{\textrm{cl}}
\newcommand{\Or}{~\textrm{or}~}
\newcommand{\Cov}{\textrm{Cov}}		
\newcommand{\Qinv}{Q_\textrm{inv}}	
\newcommand{\Ginv}{G_\textrm{inv}}
\newcommand{\Code}{{\mathcal{C}}}
\newcommand{\supp}{\textrm{supp}}		
\newcommand{\PX}{\mathcal{P}(\SrAlphabet)}
\newcommand{\PY}{\mathcal{P}(\mathcal{Z})}
\newcommand{\PXY}{\mathcal{P}(\SrAlphabet \rightarrow \mathcal{Z})}
\newcommand{\bartheta}{{\overline{\theta}}}				
\def\bf#1{\boldsymbol{#1}}
\def\vect#1{\boldsymbol{#1}}
\newcommand{\nm}{{m}}
\def\QED{\hfill$\Box$}

\newtheorem{e_theo}{Theorem}
\newtheorem{e_defin}{Definition}
\newtheorem{e_lem}{Lemma}
\newtheorem{e_prop}{Proposition}
\newtheorem{e_assump}{Assumption}
\newtheorem{e_coro}{Corollary}
\newtheorem{e_rema}{Remark}
\newtheorem{e_exam}{Example}

\maketitle

\begin{abstract}
We introduce the problem of \emph{variable-length} source resolvability, where a given target probability distribution is approximated by encoding {a} \emph{variable-length} uniform random number, and the asymptotically minimum average length {rate} of the {
uniform} random numbers, called the (variable-length) resolvability, is investigated. 
We first analyze the variable-length resolvability with the variational distance as an approximation measure. 
Next, we investigate the case under the divergence as an approximation measure. 
When the asymptotically exact approximation is required, it is shown that the resolvability under the two kinds of approximation measures coincides.
We then extend the analysis to the case of \emph{channel resolvability}, where the target distribution is the output distribution via a general channel due to the fixed general source as an input.
The obtained characterization of the channel resolvability is fully general in the sense that when the channel is just the identity mapping, the characterization reduces to the general  formula for the source resolvability.
We {also} analyze the second-order variable-length resolvability. 
\end{abstract}


\IEEEpeerreviewmaketitle

{\medskip\section{Introduction}}

Generating {a} random {number} subject to a given probability distribution has a number of applications such as in information security, statistical machine learning, and computer science.
From the viewpoint of information theory, random number generation may be considered to be a transformation (encoding) of sequences emitted from a given source, called a \emph{coin distribution}, into other sequences via a deterministic mapping \cite{Elias72,Han2003,Knuth-Yao1976}.
There have been two major types of the problems of random number generation: \emph{intrinsic randomness} \cite{Han2000b,Vembu-Verdu95} and (source) \emph{resolvability} \cite{Han-Verdu93,Steinberg-Verdu96}. 
In the former case, {a} fixed-length uniform random number {is} extracted from an arbitrary coin distribution, and we want to find the \emph{maximum} achievable rate of the uniform random numbers.
In the latter case, in contrast, {a} fixed-length uniform random number used as a coin distribution {is} encoded to approximate a given \emph{target distribution}, and we want to find the \emph{minimum} achievable rate of the uniform random numbers.
Thus, there is a duality between these two problems. 

The problem of intrinsic randomness has been extended to the case of \emph{variable-length} uniform random numbers, for which the length of the random numbers {may vary}. 
This problem, referred to as the \emph{variable-length intrinsic randomness}, was first introduced by Vembu and Verd\'u \cite{Vembu-Verdu95} for a finite source alphabet and later extended by Han \cite{Han2000b} to a countably infinite alphabet.
The use of variable-length uniform random numbers can generally increase the average length rate of the uniform random numbers compared to the fixed-length ones.
This fact raises the natural question: can we also decrease the average length rate in the resolvability problem by using variable-length uniform random numbers?
Despite the duality between the two problems of random number generation, {the}  counterpart in the resolvability problem has not {yet} been discussed yet.

In this paper, we introduce the problem of \emph{variable-length} source/channel resolvability, where a given target probability distribution is approximated by encoding {a} \emph{variable-length} uniform random number. 
Distance measures between {the}  target distribution and {the}  approximated distribution are used to measure the fineness of approximation. 
We first analyze the fundamental limit on the variable-length source resolvability with the variational distance as {an approximation measure} in Sec.\ \ref{sec:VL-resolvability-variational-dist} and \ref{sec:d-VL-resolvability-variational-dist}. 
We use the smooth R\'enyi entropy of order one \cite{Renner-Wolf2004} to characterize the $\delta$-source resolvability, which is defined as the minimum achievable length rate of uniform random numbers with the asymptotic distance less than or equal to $\delta \in [0,1)$. 
In the proof of the direct part, we will develop a simple version of \emph{information spectrum slicing} \cite{Han2003}, in which each ``sliced'' information density quantized to an integer is approximated by {a fixed-length} uniform random number.
Due to the simplicity of the method, the analysis on the average rate and the variational distance is facilitated. 
We then extend the analysis to the case under the (unnormalized) divergence as {an approximation measure} in Sec.\ \ref{sec:d-VL-resolvability-div}. 
When $\delta =0$, that is, the asymptotically exact approximation is required, it is shown that the 0-source resolvability under the two kinds of {approximation} measures coincides with each other.

In Sec.\ \ref{sec:channel_resolvability}, we then consider the problem of \emph{channel resolvability} \cite{Han-Verdu93}, in which 
not only a source but a channel is fixed and the output distribution via the channel is now the target of approximation. 
We consider two types of problems in which either a general source (\emph{mean-channel resolvability}) or {a} variable-length uniform random number (\emph{variable-length channel resolvability}) is used as a coin distribution.
It is shown that the established formulas are equal for both coin distributions.
In the special case where the channel is the identity mapping, the established formulas reduce to those in source resolvability established in Sec.\ \ref{sec:VL-resolvability-variational-dist}--Sec.\ \ref{sec:d-VL-resolvability-div}.
We also analyze the second-order fundamental limits on the variable-length {channel/source} resolvability in Sec.\ \ref{sec:second-order}. 
In {this paper}, it is shown that the {variable-length} $\delta$-\emph{source} resolvability under the variational distance is equal to the minimum achievable rate by \emph{fixed-to-variable length source codes} with the error probability less than or equal to $\delta$. 
It is demonstrated that this close relationship provides a single letter characterization for the first- and second-order source resolvability under the variational distance when the source is stationary and memoryless. 

\medskip\section{Fixed-Length Resolvability: Review} \label{sec:FL-resolvability}

Let $\SrAlphabet = \{ 1, 2, \ldots, K \}$ be a finite alphabet of size $K$, and let $\CdAlphabet$ be a finite or countably infinite alphabet.
Let ${\vect{X}} = \{ \TargetSr\}_{n =1}^\infty$ be a \emph{general source} \cite{Han2003}, where $P_{\TargetSr}$ is a probability distribution on $\CdAlphabet^n$.
We do not impose any assumptions such as stationarity or ergodicity.
{In this paper, we} identify $X^n$ with its probability distribution $P_{X^n}$, and these symbols are used interchangeably.

\begin{figure}
\vspace*{-9mm}
\begin{center}
\includegraphics[width=12.5cm]{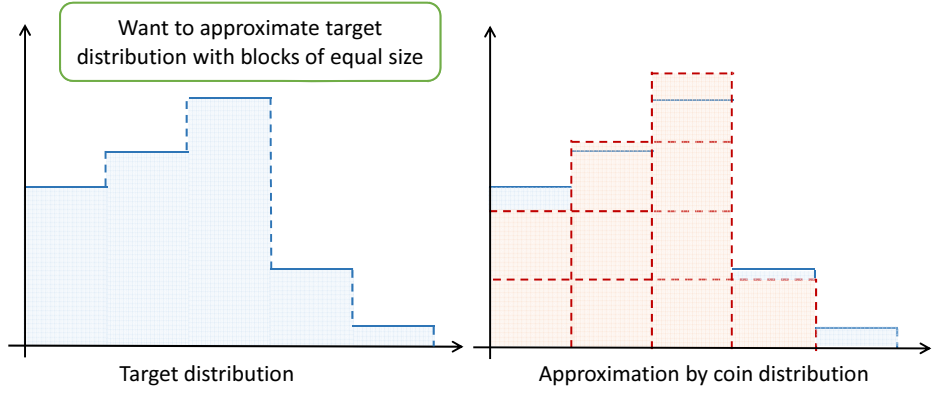}
\end{center}
\vspace*{-5mm}
\caption{Illustration of the problem of fixed-length resolvability} \label{fig:FL-resolvability}
\end{figure}

\smallskip
We {first} review the {problem of \emph{fixed-length} (source) resolvability} \cite{Han2003} using {the} variational distance as {an approximation measure}.
Let ${U_{M_n}}$ denote {the} \emph{uniform random number}, which is a random variable \emph{uniformly} distributed over ${\mathcal{U}_{M_n}} := \{ 1, \ldots, M_n\}$.
Consider {the} problem of approximating {the} \emph{target distribution} $P_{\TargetSr}$ {by} using ${U_{M_n}}$ as {the} \emph{coin distribution} via a deterministic mapping $\varphi_n : \{ 1, \ldots, M_n\} \rightarrow \CdAlphabet^n$.
Denoting $\ApproxSr = \varphi_n({U_{M_n}})$, we want to make $P_{\ApproxSr}$ approximate $P_{\TargetSr}$ (cf.\ Figure \ref{fig:FL-resolvability}). 
A standard choice of {the} performance measure of approximation is
 \begin{align}
 d(P_{\TargetSr}, P_{\ApproxSr}) := \frac{1}{2} \sum_{\TargetSeq \in \CdAlphabet^n} |P_{\TargetSr}(\TargetSeq) - P_{\ApproxSr} (\TargetSeq)|,   
 \end{align} 
which is referred to as the \emph{variational distance} between $P_{\TargetSr}$ and $P_{\ApproxSr} $.
It is easily seen {that} 
\begin{align}
0 &\le d(P_{\TargetSr}, P_{\ApproxSr}) \le 1. 
\end{align}

Let us now define the problem {for \emph{source resolvability}}.
\begin{e_defin}[Fixed-Length Resolvability]
A resolution rate $R \ge 0$ is said to be \emph{fixed-length achievable} {or simply f-\emph{achievable}} {(under {the} variational distance)} if there exists a deterministic mapping $\varphi_n : \{ 1, \ldots, M_n\} \rightarrow \CdAlphabet^n$ satisfying\footnote{Throughout this paper, logarithms are of base $K$.}
\begin{align}
 \limsup_{n \rightarrow \infty} \frac{1}{n} \log M_n &\le R, \label{eq:fixed_length_rate_cond}  \\ 
 \lim_{n \rightarrow \infty} d(P_{\TargetSr}, P_{\ApproxSr}) &= 0 \label{eq:fixed_length_variational_dist_cond},
\end{align}
where $\ApproxSr = \varphi_n({U_{M_n}})$ and ${U_{M_n}}$ is {the} uniform random number over {$\mathcal{U}_{M_n}$}.
The infimum of {f-achievable} rates:
\begin{align}
S_{\rm f} ({\vect{X}}):= \inf \{ R : ~ R ~\mbox{is f-achievable}\} \label{eq:fixed_length_opt_rate}
\end{align}
is called the \emph{fixed-length resolvability} {or simply f-\emph{resolvability}}.
\QED
\end{e_defin}

\medskip
The following result is given by Han and Verd\'u \cite{Han-Verdu93}.
\begin{e_theo}[{Han and Verd\'u \cite{Han-Verdu93}}] \label{theo:fixed_length_resolvability}
For any general target source ${\vect{X}}$,
\begin{align}
S_{\rm f} ({\vect{X}}) = \overline{H}({\vect{X}}), \label{eq:fixed_length_formula}
\end{align}
where
\begin{align}
\overline{H}({\vect{X}}) &:= \inf \left\{a : \lim_{n \rightarrow \infty} \Pr\left\{ \frac{1}{n} \log \frac{1}{P_{\TargetSr}(\TargetSr)} > a \right\} = 0 \right\}.
\end{align}
\QED
\end{e_theo}

\medskip
The following problem is called the \emph{$\delta$-resolvability problem} \cite{Han2003,Steinberg-Verdu96}, which relaxes the condition on the variational distance, compared to \eqref{eq:fixed_length_variational_dist_cond}.
\begin{e_defin}[$\delta$-Fixed-Length Resolvability]
For a fixed $\delta \in [0, 1)$, a resolution rate $R \ge 0$ is said to be \emph{$\delta$-fixed-length achievable} {or simply $\mathrm{f}(\delta)$-\emph{achievable}} {(under {the} variational distance)} if there exists a deterministic mapping $\varphi_n : \{ 1, \ldots, M_n\} \rightarrow \CdAlphabet^n$ satisfying
\begin{align}
  \limsup_{n \rightarrow \infty} \frac{1}{n} \log M_n &\le R, \label{eq:fixed_length_rate_cond2} \\
 \limsup_{n \rightarrow \infty} d(P_{\TargetSr}, P_{\ApproxSr}) &\le \delta, \label{eq:d_fixed_length_variational_dist_cond}   
\end{align}
where $\ApproxSr = \varphi_n({U_{M_n}})$ and ${U_{M_n}}$ is {the} uniform random number over {$\mathcal{U}_{M_n}$}.
The infimum of all {$\mathrm{f}(\delta)$-achievable} rates:
\begin{align}
S_{\rm f} (\delta|{\vect{X}}):= \inf \{ R : ~ R ~\mbox{is {$\mathrm{f}(\delta)$-achievable}}\} \label{eq:d_fixed_length_opt_rate}
\end{align}
is {referred to as} the \emph{$\delta$-fixed-length resolvability} or simply {$\mathrm{f}(\delta)$-\emph{resolvability}}.
\QED
\end{e_defin}

\medskip
\begin{e_theo}[{Steinberg and Verd\'u \cite{Steinberg-Verdu96}}] \label{theo:d-fixed_length_resolvability}
For any general target source ${\vect{X}}$,
\begin{align}
S_{\rm f} (\delta | {\vect{X}}) = \overline{H}_{\delta}({\vect{X}})~~~~(\delta \in [0, 1)), \label{eq:d-fixed_length_formula}
\end{align}
where
\begin{align}
\overline{H}_{\delta} ({\vect{X}}) &:= \inf \left\{a : \limsup_{n \rightarrow \infty} \Pr\left\{ \frac{1}{n} \log \frac{1}{P_{\TargetSr}(\TargetSr)} > a \right\} \le \delta \right\}.
\end{align}
\QED
\end{e_theo}

\medskip
\begin{e_rema} \label{rema:relation_source_coding}
The fixed-length resolvability problem is deeply related to the \emph{fixed-length source coding} problem allowing the probability of decoding error up to $\varepsilon$. Denoting by $R_{\rm f}(\varepsilon| {\vect{X}})$ the \emph{minimum achievable rate} for the source ${\vect{X}}$, there is the relationship {\cite{Steinberg-Verdu96}}:
\begin{align}
R_{\rm f}(\varepsilon| {\vect{X}}) = \overline{H}_{\varepsilon} ({\vect{X}})~~~(\forall \varepsilon \in [0,1))
\end{align}
and hence, by Theorem \ref{theo:d-fixed_length_resolvability}, 
\begin{align}
S_{\rm f} (\delta | {\vect{X}}) =  R_{\rm f}(\delta| {\vect{X}}) ~~~(\forall \delta \in [0,1) ). 
\end{align}
\QED
\end{e_rema}

\medskip\section{{Variable-Length} Resolvability: Variational Distance} \label{sec:VL-resolvability-variational-dist}

{
In this section, we introduce the problem of \emph{variable-length} resolvability, where {the}  target probability distribution is approximated by encoding {a} \emph{variable-length uniform random number}. 
As an initial step, we analyze the fundamental limit on the variable-length resolvability with the variational distance as {an approximation measure}.
}

\subsection{Definitions}

Let $\SrAlphabet^*$ denote the set of {all}  sequences $\CoinSeq \in \SrAlphabet^m$ {over} $m = 0, 1, 2, \cdots$, where $\SrAlphabet^0 = \{ \lambda \}$ ($\lambda$ is the null string). 
Let $L_n$ denote a random variable which takes a value in $\{0,1,2, \ldots\}$.
We define {the} \emph{variable-length uniform random number} $U^{(L_n)}$ {so} that
$U^{(m)}$ is uniformly distributed over $\SrAlphabet^m$ given $L_n = m$.
{In other words,}
\begin{align}
P_{U^{(L_n)}}(\CoinSeq, m) &:= \Pr\{ U^{(L_n)} = \CoinSeq , L_n = m \} =\frac{\Pr\{ L_n = m \}}{K^m} ~~~~(\forall \CoinSeq \in \SrAlphabet^m), \\
\Pr\{ U^{(L_n)} = \CoinSeq | L_n = m \} & = \frac{P_{U^{(L_n)}}(\CoinSeq, m) }{\Pr\{ L_n = m \}} = \frac{1}{K^m}  ~~~~(\forall \CoinSeq \in \SrAlphabet^m),
\end{align}
where $K = |\SrAlphabet|$.
It should be noticed that the variable-length sequence $\CoinSeq \in \SrAlphabet^m$ is generated with joint probability $P_{U^{(L_n)}}(\CoinSeq, m)$.

\begin{e_defin}[{Variable-Length} Resolvability: Variational Distance] \label{def:mean_achievable_rate}
A resolution rate $R \ge 0$ is said to be \emph{{variable-length achievable}} {or simply v-\emph{achievable}} (under the variational distance) if there exists a variable-length uniform random number $U^{(L_n)}$ and a deterministic mapping $\varphi_n : \SrAlphabet^* \rightarrow \CdAlphabet^n$ satisfying
\begin{align}
 \limsup_{n \rightarrow \infty} \frac{1}{n} \E [L_n] &\le R, \label{eq:VL_rate_cond}  \\ 
 \lim_{n \rightarrow \infty} d(P_{\TargetSr}, P_{\ApproxSr}) &= 0, \label{eq:VL_variational_dist_cond}   
\end{align}
where $\E[\cdot]$ denotes the expected value and $\ApproxSr = \varphi_n(U^{(L_n)})$.
The infimum of all {v-achievable} {rates}:
\begin{align}
S_{\rm v} ({\vect{X}}):= \inf \{ R : ~ R ~\mbox{is {v-achievable}}\} \label{eq:variable_length_opt_rate}
\end{align}
is called the \emph{variable-length resolvability} {or simply v-\emph{resolvability}}.
\QED
\end{e_defin}

\medskip
\begin{e_rema} \label{rema:entropy-average_length}
One may think that condition \eqref{eq:VL_rate_cond} {can} be replaced with {the condition on the sup-entropy rate}: 
\begin{align}
 \limsup_{n \rightarrow \infty} \frac{1}{n} H(U^{(L_n)}) &\le R \label{eq:VL_rate_cond_alt}
\end{align}
as in \cite{Han-Verdu93}.
{Indeed}, both conditions yield the same resolvability result.
{To see this}, let us denote by $\tilde{S}_{\rm v} ({\vect{X}})$ the infimum of {v-achievable} rates $R$ under constraints \eqref{eq:VL_variational_dist_cond}  and \eqref{eq:VL_rate_cond_alt}.
It is easily checked that
\begin{align}
\E [L_n] = \sum_{m = 1}^\infty \sum_{\CoinSeq \in \SrAlphabet^m} P_{U^{(L_n)}}(\CoinSeq, m) \log K^m &=  \sum_{m = 1}^\infty \sum_{\CoinSeq \in \SrAlphabet^m} P_{U^{(L_n)}}(\CoinSeq, m) \log \frac{\Pr\{ L_n =m \}}{P_{U^{(L_n)}}(\CoinSeq, m)} \nonumber \\
&= H(U^{(L_n)}) - H(L_n) \le H(U^{(L_n)}). \label{eq:entropy-average_length}
\end{align}
This implies $S_{\rm v} ({\vect{X}}) \le \tilde{S}_{\rm v} ({\vect{X}})$.
On the other hand, by invoking the well-known relation (cf. \cite[Corollary 3.12]{Csiszar-Korner2011}) it holds that
\begin{align}
H(L_n) \le \log (e \cdot \E [L_n]). \label{eq:length_entropy_UB}
\end{align}
Consider any resolution rate $R > S_{\rm v} ({\vect{X}})$. {Then,} \eqref{eq:VL_rate_cond} holds for some $U^{(L_n)}$ and $\varphi_n$ {and hence} \eqref{eq:length_entropy_UB} leads to
\begin{align}
\lim_{n \rightarrow \infty} \frac{1}{n}H(L_n) = \lim_{n \rightarrow \infty} \frac{1}{n} \log (e \cdot \E [L_n])= 0. \label{eq:vanishing_length_entropy}
\end{align}
{From this equation, \eqref{eq:entropy-average_length} yields} that
\begin{align}
 \limsup_{n \rightarrow \infty} \frac{1}{n} H(U^{(L_n)}) = \limsup_{n \rightarrow \infty} \frac{1}{n} \E [L_n] \le R \label{eq:entropy-length-relation}
\end{align}
{to obtain} $R \ge \tilde{S}_{\rm v} ({\vect{X}})$, implying that $S_{\rm v} ({\vect{X}}) \ge \tilde{S}_{\rm v} ({\vect{X}})$. Thus, $S_{\rm v} ({\vect{X}}) = \tilde{S}_{\rm v} ({\vect{X}})$.

{Using \eqref{eq:VL_rate_cond_alt} in place of \eqref{eq:VL_rate_cond}, the same remark applies to other resolvability problems addressed in the subsequent sections.}
\QED
\end{e_rema}

\subsection{General Formula}
In this {section}, we use the following information quantity for a general source $\vect{X}=\{ X^n\}_{n=1}^\infty$.
For  $\delta \in [0,1)$ we define
\begin{align}
G_{[\delta]} (\TargetSr) = \inf_{ \substack{\mathcal{A}_n \subseteq \CdAlphabet^n: \\ \Pr \{ \TargetSr \in \mathcal{A}_n \}  \ge 1- \delta } } \sum_{\TargetSeq \in \mathcal{A}_n}  P_{\TargetSr}(\TargetSeq) \log \frac{1}{P_{\TargetSr}(\TargetSeq)}. \label{eq:smooth_max_entropy2}
\end{align}
The $G_{[\delta]} (\TargetSr)$ is a nonincreasing function of $\delta$. 
Based on this quantity, we define 
\begin{align}
G_{[\delta]} ({\vect{X}}) &= \limsup_{n \rightarrow \infty} \frac{1}{n} G_{[\delta]} (\TargetSr) . \label{eq:d-entropy} 
\end{align}

Then, we have the following {basic} theorem:
\begin{e_theo}\label{theo:mean_resolvability}
For any general target source ${\vect{X}}$, 
\begin{align}
S_{\rm v} ({\vect{X}}) =  \lim_{\gamma \downarrow 0} G_{[\gamma]} ({\vect{X}}). \label{eq:mean_resolvability_formula}
\end{align}
\QED
\end{e_theo}
{The proof of this theorem is given below subsequently to Remark \ref{rema:weak_source_code}}.

\medskip
It has been shown by Han \cite{Han2003} that  any source ${\vect{X}} = \{ \TargetSr\}_{n = 1}^\infty$ satisfying the \emph{uniform integrability} (cf.\ Han \cite{Han2003}) satisfies
\begin{align}
\lim_{\gamma \downarrow 0} G_{[\gamma]} ({\vect{X}}) = H({\vect{X}}) := \limsup_{n \rightarrow \infty} \frac{1}{n}H(\TargetSr), \label{eq:G_and_sup-H_relation}
\end{align}
where $H({\vect{X}})$ is called \emph{sup-entropy rate}.
{Notice here, {in particular}, that the finiteness of {an} alphabet implies the uniform integrability \cite{Han2003}.}
Thus, we obtain the following corollary from Theorem \ref{theo:mean_resolvability}.

\begin{e_coro}\label{coro:mean_resolvability}
For any finite alphabet target source ${\vect{X}}$, 
\begin{align}
S_{\rm v} ({\vect{X}}) = H({\vect{X}}) . \label{eq:mean_resolvability_formula_ergodic_source}
\end{align}
\QED
\end{e_coro}

\medskip
\begin{e_rema} \label{rema:weak_source_code}
As in the case of \emph{fixed-length} resolvability and \emph{fixed-length} source coding problems,
$S_{\rm v} ({\vect{X}}) $ is tightly related to \emph{variable-length source codes} with vanishing decoding error probabilities. 
Denoting by $R_{\rm v}^*({\vect{X}})$ the \emph{minimum {error-vanishing variable-length} achievable rate} for source ${\vect{X}}$, Han \cite{Han2000} has shown that 
\begin{align}
R_{\rm v}^*({\vect{X}}) = \lim_{\gamma \downarrow 0} G_{[\gamma]} ({\vect{X}}), \label{eq:weak-source-code}
\end{align}
and hence, from Theorem \ref{theo:mean_resolvability} it is concluded that
\begin{align}
S_{\rm v} ({\vect{X}}) =  R_{\rm v}^*({\vect{X}}).
\end{align}
In addition, if a general source ${\vect{X}}$ satisfies the \emph{uniform integrability} and the strong converse property (cf.\ Han \cite{Han2003}), then equation \eqref{eq:G_and_sup-H_relation} holds and hence it {follows} from \cite[Theorem 1.7.1]{Han2003} that
\begin{align}
S_{\rm f} ({\vect{X}}) =  S_{\rm v} ({\vect{X}}) = R_{\rm v}^*({\vect{X}}) = R_{\rm v}({\vect{X}}) = R_{\rm f}({\vect{X}}) = H({\vect{X}}),
\end{align}
where $R_{\rm f}({\vect{X}}) := R_{\rm f}(0|{\vect{X}})$ and $R_{\rm v}({\vect{X}})$ denotes the minimum achievable rate of variable-length source codes with zero error probabilities for all $n = 1,2,\cdots$. 
\QED
\end{e_rema}

\medskip
\noindent
\emph{Proof of Theorem \ref{theo:mean_resolvability}}

An achievability scheme {parallels} the one {given} in the proof of Theorem \ref{theo:d-VL_resolvability} {below} with due modifications, so it is omitted {here}. 
We shall prove the converse part here. The converse part is based on the following lemma, whose proof is given in Appendix \ref{appendix:entropy_closeness_lemma}.

\begin{e_lem}\label{lem:entropy_closeness}
Let ${\vect{X}} = \{ \TargetSr\}_{n = 1}^\infty$ and $\tilde{{\vect{X}}}= \{ \ApproxSr\}_{n = 1}^\infty$ be a pair of general sources satisfying 
\begin{align}
&\lim_{n \rightarrow \infty} d(P_{\TargetSr}, P_{\ApproxSr}) = 0. \label{eq:vanishing_distance_new} 
\end{align}
Then,
\begin{align}
 \lim_{\gamma \downarrow 0} G_{[\delta+ \gamma]} ({\vect{X}}) =  \lim_{\gamma \downarrow 0} G_{[\delta+ \gamma]} (\tilde{{\vect{X}}}) ~~~(\forall \delta \in [0,1)). \label{eq:entropy_closeness}
 \end{align}
\QED
\end{e_lem}

\medskip
\quad Now let $R \ge 0$ be {variable-length achievable}. 
Then, there exists $U^{(L_n)}$ and $\varphi_n$ satisfying \eqref{eq:VL_rate_cond}  and
\eqref{eq:VL_variational_dist_cond}.
It follows from \eqref{eq:entropy-average_length} that
\begin{align}
 H(\ApproxSr) & \le H(U^{(L_n)}) = \E [L_n] +  H(L_n), \label{eq:conv_ineq5}
\end{align}
where $\ApproxSr = \varphi_n (U^{(L_n)})$ and the inequality is due to the fact that $\varphi_n$ is a deterministic mapping.
It follows from \eqref{eq:conv_ineq5} that
\begin{align}
 H(\tilde{{\vect{X}}}) = \limsup_{n \rightarrow \infty} \frac{1}{n} H(\ApproxSr) \le \limsup_{n \rightarrow \infty} \frac{1}{n} \E [L_n]  +  \limsup_{n \rightarrow \infty} \frac{1}{n} H(L_n) \le R,  \label{eq:conv_ineq7}
\end{align}
where we {have} used \eqref{eq:VL_rate_cond} and \eqref{eq:vanishing_length_entropy} for the last inequality.
Lemma \ref{lem:entropy_closeness} with $\delta = 0$ implies
\begin{align}
\lim_{\gamma \downarrow 0} G_{[\gamma]} ({\vect{X}})  &= \lim_{\gamma \downarrow 0} G_{[\gamma]} (\tilde{{\vect{X}}})  \le H(\tilde{{\vect{X}}}),
\end{align}
and thus {by \eqref{eq:conv_ineq7}}
\begin{align}
\lim_{\gamma \downarrow 0} G_{[\gamma]} ({\vect{X}})  \le R,
\end{align}
completing the proof of the converse part.
\QED

\medskip
Han and Verd\'u \cite{Han-Verdu93} {have} discussed the problem of \emph{mean-resolvability} for the target distribution $P_{\TargetSr}$ over a \emph{finite} alphabet $\CdAlphabet^n$. 
In this problem, the coin distribution may be a general source $\tilde{\vect{X}}=\{\tilde{X}^n\}_{n=1}^\infty$, where $\tilde{X}^n$ is a random variable which takes values in $\CdAlphabet^n$ {with} length rate $\frac{1}{n} \E[L_n]$  in \eqref{eq:VL_rate_cond} replaced with the entropy rate $\frac{1}{n} H(\tilde{X}^n)$.
Denoting by $\overline{S}_{\rm v}({\vect{X}})$ the mean-resolvability, which is defined as the infimum of v-achievable rates
for {a} general source ${\vect{X}}$ with \emph{countably infinite} alphabet, we can easily verify that any mean-resolution rate $R > \overline{S}_{\rm v}({\vect{X}})$ must satisfy
\begin{align}
R \ge \lim_{\gamma \downarrow 0} G_{[\gamma]}({\vect{X}})
\end{align}
by the same reasoning in the proof of the converse part of Theorem \ref{theo:mean_resolvability}.
Since $S_{\rm v} ({\vect{X}}) \ge \overline{S}_{\rm v}({\vect{X}}) $ by definition, this fact {together with Theorem \ref{theo:mean_resolvability}} indicates the following theorem, which gives a generalization of the general formula {for source resolvability} established in \cite{Han-Verdu93} for a finite alphabet $\CdAlphabet$:
\medskip
\begin{e_theo}
For any general target source ${\vect{X}}$,
\begin{align}
S_{\rm v} ({\vect{X}}) = \overline{S}_{\rm v}({\vect{X}}) = \lim_{\gamma \downarrow 0} G_{[\gamma]}({\vect{X}}). \label{eq:mean-resolvability_formula_variational_distance}
\end{align}
\QED
\end{e_theo}

\medskip\section{$\delta$-{Variable-Length} Resolvability: Variational Distance} \label{sec:d-VL-resolvability-variational-dist}

\subsection{Definitions}

{As a natural generalization of the previous preliminary section,} let us now introduce the $\delta$-resolvability problem under the variational distance using the variable-length random number, called the $\delta$-\emph{variable-length resolvability} {or simply $\mathrm{v}(\delta)$-\emph{resolvability}}.
\begin{e_defin}[$\delta$-{Variable-Length} Resolvability: Variational Distance]
A resolution rate $R \ge 0$ is said to be \emph{$\delta$-{variable-length achievable}} {or simply $\mathrm{v}(\delta)$-\emph{achievable}} {(under {the} variational distance)} {with} $\delta \in [0, 1)$ if there exists a variable-length uniform random number $U^{(L_n)}$ and
a deterministic mapping $\varphi_n : \SrAlphabet^* \rightarrow \CdAlphabet^n$ satisfying
\begin{align}
 \limsup_{n \rightarrow \infty} \frac{1}{n} \E [L_n] &\le R, \label{eq:d-variable_length_rate_cond}  \\ 
 \limsup_{n \rightarrow \infty} d(P_{\TargetSr}, P_{\ApproxSr}) &\le \delta, \label{eq:d-variable_length_variational_dist_cond}   
\end{align}
where $\ApproxSr = \varphi_n(U^{(L_n)})$.
The infimum of all {$\mathrm{v}(\delta)$-achievable} rates:
\begin{align}
S_{\rm v} (\delta| {\vect{X}}):= \inf\{ R : ~ R ~\mbox{is {$\mathrm{v}(\delta)$-achievable}}\} \label{eq:d-variable_length_opt_rate}
\end{align}
is {referred to as} the \emph{$\delta$-variable-length resolvability} {or simply $\mathrm{v}(\delta)$-resolvability}.
\QED
\end{e_defin}

\subsection{Smooth R\'enyi Entropy of Order One}

\medskip
To establish a general formula for $S_{\rm v} (\delta | {\vect{X}})$, we introduce the following quantity {for} a general source ${\vect{X}}$.
Let $\mathcal{P}(\CdAlphabet^n)$ denote the set of all probability distributions on $\CdAlphabet^n$.
For $\delta \in [0, 1)$, defining the \emph{$\delta$-ball} using the variational distance as
\begin{align}
B_{\delta}(\TargetSr) = \left\{ P_{V^n} \in \mathcal{P}(\CdAlphabet^n) : d(P_{\TargetSr}, P_{V^n}) \le \delta \right\},
\end{align}
we introduce the \emph{smooth R\'enyi entropy} of order one:
\begin{align}
H_{[\delta]} (\TargetSr) &:= \inf_{ P_{V^n} \in B_{\delta}(\TargetSr) } \sum_{\TargetSeq \in \CdAlphabet^n}  P_{V^n}(\TargetSeq) \log \frac{1}{P_{V^n}(\TargetSeq)} \nonumber \\
& = \inf_{ P_{V^n} \in B_{\delta}(\TargetSr) }  H(V^n), \label{eq:smooth_reny_entropy}
\end{align}
where $H(V^n)$ denotes the Shannon entropy of $P_{V^n}$.
The $H_{[\delta]} (\TargetSr)$ is a nonincreasing function of $\delta$. 
Based on this quantity {for} a general source ${\vect{X}} = \{ \TargetSr\}_{n =1}^\infty$, we define 
\begin{align}
H_{[\delta]} ({\vect{X}}) &= \limsup_{n \rightarrow \infty} \frac{1}{n} H_{[\delta]} (\TargetSr) . \label{eq:smooth_entropy} 
\end{align}

\begin{e_rema}
Renner and Wolf \cite{Renner-Wolf2004} {have} defined {the} \emph{smooth R\'eny entropy of order $\alpha \in (0, 1) \cup (1, \infty)$} as
\begin{align}
H_{[\delta]}^\alpha (\TargetSr) =  \inf_{ P_{V^n} \in B_{\delta}(\TargetSr) }  \frac{1}{1-\alpha} \log  \sum_{\TargetSeq \in \CdAlphabet^n} P_{V^n}(\TargetSeq)^\alpha.
\end{align}
By letting {$\alpha \uparrow 1$}, we have
\begin{align}
{\lim_{\alpha \uparrow 1 }} H_{[\delta]}^\alpha (\TargetSr) = H_{[\delta]} (\TargetSr).  \label{eq:smooth_entropy_equivalence}
\end{align}
As for the proof, see Appendix \ref{append:smooth_entropy_equivalence}.
\QED
\end{e_rema}

\subsection{General Formula}

{The following theorem indicates that the {$\mathrm{v}(\delta)$-resolvability} $S_{\rm v} (\delta|{\vect{X}})$ can be characterized by the smooth R\'enyi entropy of order one for $\vect{X}$.}
\begin{e_theo}\label{theo:d-VL_resolvability}
For any general target source ${\vect{X}}$,
\begin{align}
S_{\rm v} (\delta|{\vect{X}}) = \lim_{\gamma \downarrow 0} H_{[\delta+\gamma]} ({\vect{X}}) ~~~(\delta \in [0, 1)). \label{eq:d-mean_resolvability_formula}
\end{align}
\QED
\end{e_theo}

\begin{e_rema}
In formula \eqref{eq:d-mean_resolvability_formula}, the limit $\lim_{\gamma \downarrow 0} $ of {the} offset term $+ \gamma$ appears in the characterization of $S_{\rm v} (\delta|{\vect{X}}) $.
This is because the smooth entropy $H_{[\delta]} (X^n)$ for $X^n$ involves {the infimum over} the \emph{nonasymptotic} $\delta$-ball $B_\delta (X^n)$ for a given length $n$.
Alternatively, we may consider the \emph{asymptotic} $\delta$-ball defined as
\begin{align}
B_\delta (\vect{X}) = \left\{ \vect{V} = \{V^n \}_{n=1}^\infty: \limsup_{n \rightarrow \infty} d(P_{X^n}, P_{V^n}) \le \delta \right\}, 
\end{align} 
and then we obtain {the} alternative formula 
\begin{align}
S_{\rm v} (\delta|{\vect{X}}) = \inf_{\vect{V} \in B_\delta(\vect{X})} H (\vect{V}) ~~~(\delta \in [0, 1)) \label{eq:d-mean_resolvability_formula_alt}
\end{align}
without an offset term, where 
\begin{align}
H (\vect{V}) := \limsup_{n \rightarrow \infty} \frac{1}{n} H(V^n)
\end{align}
is the \emph{sup-entropy rate} for $\vect{V}$. 
The proof of \eqref{eq:d-mean_resolvability_formula_alt} is given in {Appendix \ref{append:formula_alt}}.
The same {remark applies to} other general formulas established in the subsequent sections.
\QED
\end{e_rema}

\medskip
\begin{e_rema}
Independently of this work, Tomita, Uyematsu, and Matsumoto \cite{TUM2016} recently have investigated the following problem: the coin distribution is given by \emph{fair coin tossing} and the average number of coin tosses should be asymptotically minimized while the variational distance between the target and approximated distributions should satisfy {\eqref{eq:d-variable_length_variational_dist_cond}}.
In this case, the asymptotically minimum average number of coin tosses is also characterized by the right-hand side {(r.h.s.)} of \eqref{eq:d-mean_resolvability_formula} (cf.\ \cite{TUM2016}).
Since the coin distribution is restricted to the one given by fair coin tossing with a stopping algorithm, the realizations of $L_n$ must satisfy the Kraft inequality (prefix codes),
whereas the problem addressed in this paper allows the probability distribution of $L_n$ to be an arbitrary discrete distribution, not necessarily implying prefix codes.
In this sense, our problem is more relaxed, while out coin is constrained to be conditionally independent given $L_n$.
Theorem \ref{theo:d-VL_resolvability} indicates that the $\mathrm{v}(\delta)$-resolvability does not change in both problems.  
Later, we shall show that even in the case where the coin distribution may be any general source $\vect{X}$, the  $\delta$-resolvability remains still the same (cf.\ Theorem \ref{theo:d-VL_ch_resolvability} and Remark \ref{rema:source_resolvability_recapture}).
\QED    
\end{e_rema}

\medskip
\begin{e_rema}
{Analogously} to the case $\delta = 0$, there is a deep relation between this $\delta$-resolvability problem and $\delta$-variable-length source coding {with error probability asymptotically not exceeding $\delta$}.
Han \cite{Han2000} and Koga and Yamamoto \cite{Koga-Yamamoto2005} have shown that the minimum average length rate $R_{\rm v}^*(\delta|{\vect{X}})$ of $\delta$-variable-length source codes is {given by}
\begin{align}
R_{\rm v}^* (\delta|{\vect{X}}) = \lim_{\gamma \downarrow 0} G_{[\delta+\gamma]} ({\vect{X}}) ~~~(\forall \delta \in [0,1)). \label{eq:d-source_coding_formula}
\end{align}
{Theorem \ref{theo:d-VL_resolvability} and Proposition \ref{prop:entropy_relation} to be shown below reveal that 
\begin{align}
S_{\rm v} (\delta|{\vect{X}})  = R_{\rm v}^* (\delta|{\vect{X}})  ~~~(\forall \delta \in [0,1)).  
\end{align}
} 
\QED
\end{e_rema}

\medskip
{The following proposition shows a general relationship between $G_{[\delta]}({\vect{X}})$ and $H_{[\delta]}({\vect{X}})$.}
\begin{e_prop}\label{prop:entropy_relation}
For any general source ${\vect{X}}$,
\begin{align}
H_{[\delta]} ({\vect{X}}) =   G_{[\delta]} ({\vect{X}}) \le (1-\delta) \overline{H}_\delta ({\vect{X}}) ~~~(\forall \delta \in [0, 1)). \label{eq:asymptotic_equivalence2}
\end{align}
In particular,
\begin{align}
\lim_{\gamma \downarrow 0 }H_{[\delta+\gamma]} ({\vect{X}}) = \lim_{\gamma \downarrow 0} G_{[\delta+\gamma]} ({\vect{X}}) \le (1-\delta) \overline{H}_\delta ({\vect{X}}) ~~~(\forall \delta \in [0, 1)). \label{eq:asymptotic_equivalence3}
\end{align}
\end{e_prop}
(\emph{Proof})~~See Appendix \ref{appendix:entropy_relation}.
\QED

\medskip

{In view of Theorems \ref{theo:d-fixed_length_resolvability} and \ref{theo:d-VL_resolvability}, Proposition \ref{prop:entropy_relation} implies that we have $S_{\rm v}(\delta | \vect{X}) \le (1-\delta) S_{\rm f}(\delta | \vect{X})$ for all $\delta \in [0,1)$, where $S_{\rm f}(\delta | \vect{X})$ denotes the $\mathrm{f}(\delta)$-resolvability.
This relationship elucidates the significance of the use of variable-length uniform random numbers to minimize the average length rate. 
The proposition also demonstrates that $G_{[\delta]} ({\vect{X}})$ coincides with $H_{[\delta]} ({\vect{X}})$ for all $\delta \in [0, 1)$ for any general source $\vect{X}$.} 

\medskip
\begin{e_exam}[i.i.d.\ source] \label{exam:two_entropies}
Let ${\vect{X}} = \{ X^n\}_{n=1}^\infty$ be {a} source with i.i.d.\ {$X^n = (X_1, X_2, \ldots, X_n)$}.
By the weak law of large numbers, $\frac{1}{n} \log \frac{1}{P_{X^n} (X^n)}$ ({the} sum of independent random variables) converges to the Shannon entropy $H(X_1)$ in probability.
Therefore, 
\begin{align}
S_{\rm f} ({\delta}|{\vect{X}}) = \overline{H}_{{\delta}}({\vect{X}}) = H(X_1) ~~~(\forall \delta \in [0,1)). \nonumber
\end{align}
On the other hand, it holds that
\begin{align}
S_{\rm v} (\delta|{\vect{X}})  = H_{[\delta]} (\vect{X}) = G_{[\delta]} (\vect{X}) = (1 - \delta) H(X_1) ~~~(\forall \delta \in [0,1)), 
\end{align}
where the second and third equalities are due to Proposition \ref{prop:entropy_relation} and \cite[Appendix III]{Koga-Yamamoto2005}, respectively.
Thus, we have $S_{\rm v} (\delta|{\vect{X}}) \le S_{\rm f} (\delta|{\vect{X}})$, where the inequality is strict for $\delta \in (0,1)$ if $H(X_1) > 0$.
\QED
\end{e_exam}

\medskip
\noindent
\emph{Proof of Theorem \ref{theo:d-VL_resolvability}}

\emph{1) Converse Part:}~~~ Let $R$ be {$\mathrm{v}(\delta)$-achievable}. 
Then, there exists $U^{(L_n)}$ and $\varphi_n$ satisfying \eqref{eq:d-variable_length_rate_cond}  and
\begin{align}
 &\limsup_{n \rightarrow \infty} \delta_n \le \delta, \label{eq:d-variable_length_variational_dist_cond2}   
\end{align}
where we define $\delta_n = d(P_{\TargetSr}, P_{\ApproxSr})$ with $\ApproxSr = \varphi_n (U^{(L_n)})$.
Equation \eqref{eq:d-variable_length_variational_dist_cond2} implies that for any given $\gamma >0$, $\delta_n \le \delta + \gamma$ for all {$n \ge n_0$ with some $n_0 >0$}, and therefore
\begin{align}
H_{[\delta + \gamma]}(\TargetSr) \le H_{[\delta_n]}(\TargetSr)~~~(\forall n \ge n_0), \label{eq:conv_ineq1}
\end{align}
because $H_{[\delta]}(\TargetSr)$ is a nonincreasing function of $\delta$.
Since $P_{\ApproxSr} \in B_{\delta_n}(\TargetSr)$, we have
\begin{align}
H_{[\delta_n]}(\TargetSr) \le H(\ApproxSr). \label{eq:conv_ineq2}
\end{align}
On the other hand, it follows from \eqref{eq:entropy-average_length} that
\begin{align}
 H(\ApproxSr) & \le H(U^{(L_n)}) = \E [L_n] +  H(L_n), \label{eq:conv_ineq3}
\end{align}
where the inequality is due to the fact that $\varphi_n$ is a deterministic mapping and $\ApproxSr = \varphi_n (U^{(L_n)})$.

Combining  \eqref{eq:conv_ineq1}--\eqref{eq:conv_ineq3} yields
\begin{align}
H_{[\delta+\gamma]} ({\vect{X}}) &= \limsup_{n \rightarrow \infty} \frac{1}{n} H_{[\delta + \gamma]}(\TargetSr) \nonumber \\
&\le \limsup_{n \rightarrow \infty} \frac{1}{n} \E [L_n]  +  \limsup_{n \rightarrow \infty} \frac{1}{n} H(L_n) \le R, 
\end{align}
where we {have used} \eqref{eq:vanishing_length_entropy} and \eqref{eq:d-variable_length_rate_cond} for the last inequality.
Since $\gamma > 0$ is arbitrary, we obtain 
\begin{align}
\lim_{\gamma \downarrow 0} H_{[\delta+\gamma]} ({\vect{X}})  \le R.
\end{align}

\QED

\emph{2) Direct Part:}~~~Without loss of generality, we assume that $H^* := \lim_{\gamma \downarrow 0} H_{[\delta + \gamma]}({\vect{X}})$ is finite ($H^* < + \infty$).
Letting $R = H^* + 3 \gamma$, where $\gamma >0$ is an arbitrary constant, we shall show that $R$ is {$\mathrm{v}(\delta)$-achievable}.
{In what follows, we use a simpler form of \emph{information spectrum slicing} \cite{Han2003}, where each sliced information quantized to a positive integer $\ell$ is approximated by the uniform random number $U^{(\ell)}$ of length $\ell$. }

First, we note that 
\begin{align}
H^* \ge H_{[\delta + \gamma]} ({\vect{X}}) \ge \frac{1}{n}  H_{[\delta + \gamma]} (\TargetSr) -\gamma~~~~(\forall n > n_0) \label{eq:H_ineq}
\end{align}
because of the monotonicity of $H_{[\delta]} ({\vect{X}})$ in $\delta$.
Let ${V^n}$ be a random variable subject to $P_{V^n} \in B_{\delta + \gamma} (\TargetSr)$ {which satisfies}
\begin{align}
H_{[\delta + \gamma]}(\TargetSr) + \gamma \ge H(V^n).  \label{eq:H_ineq2}
\end{align}
For $\gamma > 0$, we {can} choose $c_n > 0$ {so large} that
\begin{align}
\Pr \{ V^n \not\in T_n \} \le \gamma \label{eq:dominant_set}
\end{align}
where 
\begin{align}
T_n := \left\{ \TargetSeq \in \CdAlphabet^n : \frac{1}{n} \log \frac{1}{P_{V^n}(\TargetSeq)} \le c_n \right\}.
\end{align}
We also define
\begin{align}
\ell (\TargetSeq) := \left\{
\begin{array}{ll}
 \lceil \log \frac{1}{P_{V^n}(\TargetSeq)} + n \gamma \rceil & \mbox{for}~ \TargetSeq \in T_n \\
 0 & \mbox{otherwise}. 
\end{array} 
\right. 
\end{align}
Letting for $m=0, 1, \ldots, \beta_n:=  \lceil n (c_n+\gamma) \rceil$
\begin{align}
S_n(m) := \left\{ \TargetSeq \in \CdAlphabet^n : \ell (\TargetSeq) = m\right\},
\end{align}
these sets form a partition of $\CdAlphabet^n$. i.e.,
\begin{align}
\bigcup_{m = 0}^{\beta_n} S_n(m) = \CdAlphabet^n~~~\mbox{and}~~~\bigcup_{m = 1}^{\beta_n} S_n(m) = T_n. \label{eq:partition}
\end{align}
We set $L_n$ {so} that
\begin{align}
\Pr \{ L_n = m\} = \Pr\{V^n \in S_n(m)\},
\end{align}
{where it is obvious that $\sum_{m = 0}^{\beta_n} \Pr \{ L_n = m\} = 1$}, 
and hence the probability distribution of {the} variable-length uniform random number $U^{(L_n)}$ is
\begin{align}
P_{U^{(L_n)}}(\CoinSeq, m) := \Pr \{ U^{(L_n)} = \CoinSeq, L_n = m \} = \frac{\Pr\{V^n \in S_n(m)\}}{K^m}~~(\forall \CoinSeq \in \SrAlphabet^m).
\end{align}

\medskip
\noindent \emph{Construction of Mapping $\varphi_n: \SrAlphabet^* \rightarrow \CdAlphabet^n$:} \\
%
\quad 
Index the elements in $S_n(m)$ as $\TargetSeq_1, \TargetSeq_2, \ldots,\TargetSeq_{|S_n(m)|}~{(m=1,2,\cdots)}$, where 
\begin{align}
 |S_n(m)| \le K^{m - n\gamma} \label{eq:size_Sn(m)}
 \end{align}
  since for $\TargetSeq \in S_n(m)$
\begin{align}
\log \frac{1}{P_{V^n}(\TargetSeq)} \le m -  n \gamma ~\Longleftrightarrow ~ P_{V^n}(\TargetSeq)  \ge K^{-(m - n\gamma)}, \label{eq:set_Sn(m)}
\end{align}
and therefore
\begin{align}
1 \ge \sum_{\TargetSeq \in S_n(m)} P_{V^n}(\TargetSeq) \ge \sum_{\TargetSeq \in S_n(m)} K^{-(m - n\gamma)} = |S_n(m)| K^{-(m - n\gamma)}. \label{eq:size_eval}
\end{align}
For $i = 1, 2,\ldots, |S_n(m)|$, define $\tilde{A}_i^{(m)} \subset \SrAlphabet^m$ as {the} set of sequences $\CoinSeq \in \SrAlphabet^m$ {so} that
\begin{align}
\sum_{\CoinSeq \in \tilde{A}_i^{(m)}} P_{U^{(L_n)}}(\CoinSeq, m) \le P_{V^n}(\TargetSeq_i) < \sum_{\CoinSeq \in \tilde{A}_i^{(m)}} P_{U^{(L_n)}}(\CoinSeq, m)  + \frac{\Pr\{V^n \in S_n(m)\}}{K^m} \label{eq:prob_set1}
\end{align}
and
\begin{align}
\tilde{A}_i^{(m)} \cap \tilde{A}_j^{(m)} = \emptyset~~~~(i \neq j). \label{eq:prob_set1b}
\end{align}
If
\begin{align}
\sum_{i =1}^{|S_n(m)|} \sum_{\CoinSeq \in \tilde{A}_i^{(m)}}  P_{U^{(L_n)}}(\CoinSeq, m) <  \sum_{i =1}^{|S_n(m)|}  P_{V^n}(\TargetSeq_i) = \Pr \{ V^n \in S_n(m)\},
\end{align}
then add {a} $\CoinSeq_i \in \SrAlphabet^m \setminus (\cup_j \tilde{A}_j^{(m)})$ to {obtain}
\begin{align}
A_i^{(m)} = \tilde{A}_i^{(m)} \cup \{ \CoinSeq_i \}
\end{align} 
for $i = 1, 2, \ldots $  in order, 
until it holds that with some $ 1\le c  \le |S_n(m)|$
\begin{align}
\bigcup_{i =1}^{c} A_i^{(m)}\cup \bigcup_{i =c+1}^{|S_n(m)|} \tilde{A}_i^{(m)} = \SrAlphabet^m,
\end{align}
{where $\CoinSeq_1, \CoinSeq_2, \cdots$ are {selected} to be all different.}
Since $|\SrAlphabet^m| = K^m$ and
\begin{align}
\sum_{\CoinSeq \in \SrAlphabet^m} P_{U^{(L_n)}}(\CoinSeq, m) = \sum_{\CoinSeq \in \SrAlphabet^m} \frac{\Pr\{ V^n \in S_n(m)\}}{K^m} = \Pr\{ V^n \in S_n(m)\},
\end{align}
such $1 \le c \le |S_n(m)|$ always exists.
For simplicity, we set for $i = c+1, {c+2}, \ldots, |S_n(m)|$ 
\begin{align}
A_i^{(m)} = \tilde{A}_i^{(m)}
\end{align}
and for $i = 1, {2}, \ldots, |S_n(m)|$ 
\begin{align}
\varphi_n (\CoinSeq) = \TargetSeq_i ~~\mbox{for}~\CoinSeq \in A_i^{(m)},
\end{align}
{which defines the random variable $\tilde{X}^n$ with values in $\mathcal{X}^n$ such that}
\begin{align}
P_{\ApproxSr}(\TargetSeq_i) = \sum_{\CoinSeq \in A_i^{(m)}}  P_{U^{(L_n)}}(\CoinSeq, m)~~~{(\TargetSeq_i \in S_n(m))}, \label{eq:Prob_tildeY}
\end{align}
that is, $\tilde{X}^n=\varphi_n(U^{(L_n)})$,
where if $\CdAlphabet^n \setminus T_n \neq \emptyset$, we choose some $\TargetSeq_0 \in \CdAlphabet^n \setminus T_n$ and set
\begin{align}
P_{\ApproxSr}(\TargetSeq_0) = \Pr \{ V^n \not\in T_n \}~~\mathrm{and}~~\varphi_n (\lambda) = \TargetSeq_0. \label{eq:Prob_tildeY2}
\end{align}
Notice that by this construction
we have
\begin{align}
 |P_{\ApproxSr}(\TargetSeq_i) - P_{V^n}(\TargetSeq_i)| \le \frac{\Pr \{ V^n \in S_n(m)\}}{K^m} \label{eq:prob_set2}
\end{align}
for $i = 1, 2, \ldots, |S_n(m)|; m = 1, 2, \ldots, \beta_n$, and
\begin{align}
\Pr \{ \ApproxSr \not\in T_n \} = \Pr \{ V^n \not\in T_n \} \le \gamma. \label{eq:Prob_tildeY3}
\end{align}

\medskip
\noindent \emph{Evaluation of Average Length:} \\
\quad
The average length $\E [L_n]$ is evaluated as follows:
\begin{align}
\E[L_n] &= \sum_{m=1}^{\beta_n}  \sum_{\CoinSeq \in \SrAlphabet^m} P_{U^{(L_n)}}(\CoinSeq, m) \cdot m \nonumber \\
&=\sum_{m=1}^{\beta_n} \sum_{i=1}^{|S_n(m)|} \sum_{\CoinSeq \in A_i^{(m)}} P_{U^{(L_n)}}(\CoinSeq, m) \cdot m \nonumber \\
&= \sum_{m=1}^{\beta_n} \sum_{\TargetSeq_i \in S_n(m)} P_{\ApproxSr} (\TargetSeq_i) \cdot m, \label{eq:average_length1a}
\end{align}
where we have used $\SrAlphabet^m = \bigcup_{i=1}^{|S_n(m)|} A_i^{(m)}$ and \eqref{eq:Prob_tildeY}.
For $\TargetSeq_i \in S_n(m)$ we obtain from \eqref{eq:prob_set2}
\begin{align}
P_{\ApproxSr}(\TargetSeq_i) &\le P_{V^n}(\TargetSeq_i) + \frac{\Pr\{ V^n \in S_n(m)\}}{K^m}  \nonumber \\
 & \le  P_{V^n}(\TargetSeq_i) \left( 1 + \frac{1}{P_{V^n}(\TargetSeq_i)K^m} \right) \nonumber \\
 & \le P_{V^n}(\TargetSeq_i) \left( 1 + \frac{1}{K^{n\gamma}} \right), \label{eq:bound_P_tildeY}
\end{align}
where to derive the last inequality we have used \eqref{eq:set_Sn(m)}.
Plugging the inequality
\begin{align}
m \le \log \frac{1}{P_{V^n}(\TargetSeq_i)} + n\gamma + 1 ~~(\forall \TargetSeq_i \in S_n(m))
\end{align}
and \eqref{eq:bound_P_tildeY} into \eqref{eq:average_length1a}, we obtain
\begin{align}
\E[L_n]  &\le \left( 1 + \frac{1}{K^{n\gamma}} \right) \sum_{m=1}^{\beta_n} \sum_{\TargetSeq_i \in S_n(m)} P_{V^n}(\TargetSeq_i) \left(\log \frac{1}{P_{V^n}(\TargetSeq_i)} + n\gamma + 1\right) \nonumber \\
&\le \left( 1 + \frac{1}{K^{n\gamma}} \right) \left( H(V^n) + n\gamma + 1 \right), \label{eq:average_length}
\end{align}
{which yields}
\begin{align}
\limsup_{n \rightarrow \infty} \frac{1}{n} \E [L_n] &\le \limsup_{n \rightarrow \infty} \frac{1}{n} H(V^n) + 2 \gamma \nonumber \\
&\le  \limsup_{n \rightarrow \infty} \frac{1}{n} H_{[\delta + \gamma]} (\TargetSr) + 3 \gamma \nonumber \\
&=  H_{[\delta + \gamma]} ({\vect{X}}) + 3 \gamma \nonumber \\
&\le H^* + 3 \gamma = R, \label{eq:rate_cond_satisfy}
\end{align}
where the second inequality follows from \eqref{eq:H_ineq2} and the last one is due to \eqref{eq:H_ineq}.

\medskip
\noindent \emph{Evaluation of Variational Distance:} \\
\quad 
From \eqref{eq:size_Sn(m)} and \eqref{eq:prob_set2} we have
\begin{align}
\sum_{\TargetSeq \in S_n(m)} |P_{\ApproxSr}(\TargetSeq) - P_{V^n}(\TargetSeq)| \le  \frac{|S_n(m)| \Pr\{V^n \in S_n(m)\}}{K^m} \le \frac{\Pr\{V^n \in S_n(m)\}}{K^{n\gamma}},
\end{align}
leading from {\eqref{eq:partition}} to
\begin{align}
d(P_{\ApproxSr}, P_{V^n}) &= \frac{1}{2} \sum_{\TargetSeq \in T_n} |P_{\ApproxSr}(\TargetSeq) - P_{V^n}(\TargetSeq)| + \frac{1}{2} \sum_{\TargetSeq \not\in T_n} |P_{\ApproxSr}(\TargetSeq) - P_{V^n}(\TargetSeq)|  \nonumber \\
&\le \frac{1}{2} \sum_{m=1}^{\beta_n} \sum_{\TargetSeq \in S_n(m)} |P_{\ApproxSr}(\TargetSeq) - P_{V^n}(\TargetSeq)| + \frac{1}{2} \left( \Pr \{ \ApproxSr \not\in T_n\} + \Pr \{ V^n \not\in T_n\} \right) \nonumber \\
&\le \frac{1}{2} \sum_{m=1}^{\beta_n} \frac{\Pr\{V^n \in S_n(m)\}}{K^{n\gamma}} +  \gamma \le \frac{1}{2}  K^{-n\gamma} + \gamma, \label{eq:distance_UB1}
\end{align}
where we have used \eqref{eq:Prob_tildeY3} to obtain the leftmost inequality in \eqref{eq:distance_UB1}.
By the triangle inequality, we obtain
\begin{align}
d(P_{\TargetSr}, P_{\ApproxSr}) \le d(P_{\TargetSr}, P_{V^n}) + d(P_{\ApproxSr}, P_{V^n}) \le \delta + 2 \gamma + \frac{1}{2} K^{-n\gamma}, \label{eq:distance_UB}
\end{align}
where the last inequality follows because $P_{V^n} \in B_{\delta + \gamma} (\TargetSr)$.
Thus, we obtain from \eqref{eq:distance_UB} 
\begin{align}
\limsup_{n \rightarrow \infty} d(P_{\TargetSr}, P_{\ApproxSr})  &\le \delta + 2\gamma.
\end{align}
Since $\gamma > 0$ is arbitrary and we have \eqref{eq:rate_cond_satisfy}, $R$ is {$\mathrm{v}(\delta)$-achievable}.
\QED

\medskip\section{$\delta$-Variable-Length Resolvability: Divergence} \label{sec:d-VL-resolvability-div}

So far, we have considered the problem of the variable-length resolvability, in which the approximation level is measured by the variational distance between $X^n$ and $\tilde{X}^n$.
It is sometimes of use to deal with other quantity {as an approximation measure}.
In this section, we use the (unnormalized) divergence as {the approximation measure}. 

\subsection{Definitions}

In this section, we address the following problem.
\begin{e_defin}[$\delta$-{Variable-Length} Resolvability: Divergence]
A resolution rate $R \ge 0$ is said to be \emph{$\delta$-{variable-length achievable}} {or simply $\mathrm{v}_D(\delta)$-achievable} (under the divergence) {with} $\delta \ge 0$ if there exists a variable-length uniform random number $U^{(L_n)}$ and
a deterministic mapping $\varphi_n : \SrAlphabet^* \rightarrow \CdAlphabet^n$ satisfying
\begin{align}
\limsup_{n \rightarrow \infty} \frac{1}{n} \E [L_n] &\le R,  \label{eq:d-variable_length_rate_cond_divergence} \\
 \limsup_{n \rightarrow \infty} D(\ApproxSr || \TargetSr ) &\le \delta, \label{eq:d-variable_length_divergence_cond}   
\end{align}
where $\ApproxSr = \varphi_n(U^{(L_n)})$ and $ D(\ApproxSr || \TargetSr ) $ denotes {the} \emph{divergence} between $P_{\ApproxSr}$ and $P_{\TargetSr}$ defined by
\begin{align}
 D(\ApproxSr || \TargetSr ) = \sum_{\TargetSeq \in \CdAlphabet^n} P_{\ApproxSr}(\TargetSeq) \log \frac{P_{\ApproxSr}(\TargetSeq)}{P_{\TargetSr}(\TargetSeq)}.
\end{align}
The infimum of all {$\mathrm{v}_D(\delta)$-achievable} rates:
\begin{align}
S_{\rm v}^D (\delta| {\vect{X}}):= \inf\{ R : ~ R ~\mbox{is {$\mathrm{v}_D(\delta)$-achievable}} \} \label{eq:d-variable_length_opt_rate_divergence}
\end{align}
is called the \emph{$\delta$-variable-length resolvability} {or simply $\mathrm{v}_D(\delta)$-resolvability}.
\QED
\end{e_defin}

\medskip
To establish {the} general formula for $S_{\rm v}^D (\delta | {\vect{X}})$, we introduce the following quantity {for} a general source ${\vect{X}}=\{ X^n\}_{n=1}^\infty$.
{Recall that $\mathcal{P}(\CdAlphabet^n)$ denotes} the set of all probability distributions on $\CdAlphabet^n$.
For $\delta \ge 0$, defining the \emph{$\delta$-ball} using {the} divergence as
\begin{align}
B_{\delta}^D(\TargetSr) = \left\{ P_{V^n} \in \mathcal{P}(\CdAlphabet^n) : D(V^n||\TargetSr) \le \delta \right\},
\end{align}
we introduce the following quantity, referred to as {the} \emph{smooth entropy} using {the} divergence:
\begin{align}
H_{[\delta]}^D (\TargetSr) &:= \inf_{ P_{V^n} \in B_{\delta}^D(\TargetSr) } H(V^n), \label{eq:smooth_divergence_entropy}
\end{align}
where $H(V^n)$ denotes the Shannon entropy of $P_{V^n}$.
The $H_{[\delta]}^D (\TargetSr)$ is a nonincreasing function of $\delta$. 
Based on this quantity, for a general source ${\vect{X}} =\{ \TargetSr \}_{n=1}^\infty$ we define 
\begin{align}
H_{[\delta]}^D ({\vect{X}}) &= \limsup_{n \rightarrow \infty} \frac{1}{n} H_{[\delta]}^D (\TargetSr) . \label{eq:smooth_divergence_entropy2} 
\end{align}

\medskip
The following lemma is used to derive Corollary \ref{coro:mean_resolvability_relation} of Theorem \ref{theo:d-VL_resolvability_div} below in the next subsection.
\begin{e_lem}\label{lem:entropy_relation_divergence}
For any general source ${\vect{X}}$,
\begin{align}
H_{[\delta]} ({\vect{X}}) \le  H_{[g(\delta)]}^D ({\vect{X}})  ~~~( \delta {\, \ge 0}), \label{eq:asymptotic_equivalence4}
\end{align}
where we define ${g(\delta) = 2 \delta^2/  \ln K}$, and 
\begin{align}
\lim_{\delta \downarrow 0} G_{[\delta]} ({\vect{X}}) = \lim_{\delta \downarrow 0 }H_{[\delta]} ({\vect{X}}) =  \lim_{\delta \downarrow 0 }H_{[\delta]}^D ({\vect{X}}) \le \overline{H} ({\vect{X}}). \label{eq:asymptotic_equivalence5}
\end{align}
(\emph{Proof}) See Appendix \ref{appendix:proof_entropy_relation_divergence}.
\QED
\end{e_lem}

\subsection{General Formula}

We establish the following theorem which characterizes $S_{\rm v}^D (\delta|{\vect{X}}) $ for all $\delta \ge 0$ {in terms of} the smooth entropy using the divergence.
\begin{e_theo}\label{theo:d-VL_resolvability_div}
For any general target source ${\vect{X}}$,
\begin{align}
S_{\rm v}^D (\delta|{\vect{X}}) = \lim_{\gamma \downarrow 0} H_{[\delta+\gamma]}^D ({\vect{X}}) ~~~( \delta \, {\ge 0}). \label{eq:d-VL_resolvability_formula_div}
\end{align}
\QED
\end{e_theo}

\medskip
\begin{e_rema} \label{rema:normalized_divergence}
It should be noticed that the {approximation measure} considered here is \emph{not} the \emph{normalized} divergence
\begin{align}
\frac{1}{n} D(\varphi_n(U^{(L_n)}) || \TargetSr),
\end{align}
which has been used in the problem of $\delta$-fixed-length resolvability \cite{Steinberg-Verdu96}.
The achievability scheme given in the proof of the direct part of Theorem \ref{theo:d-VL_resolvability_div} can also be used in the case of this relaxed measure.
Indeed, denoting the {variable-length $\delta$-resolvability} using the normalized divergence by $\tilde{S}_{\rm v}^D (\delta|{\vect{X}})$, the general formula of $\tilde{S}_{\rm v}^D (\delta|{\vect{X}})$ is given in the same form as \eqref{eq:d-VL_resolvability_formula_div}, if the {radius} of the $\delta$-ball $B_{\delta}^D(\TargetSr)$ in the definition of $H_{[\delta]}^D (\TargetSr)$ is replaced with the normalized divergence. 
It generally holds $S_{\rm v}^D (\delta|{\vect{X}}) \ge \tilde{S}_{\rm v}^D (\delta|{\vect{X}})$ for all $\delta \ge 0$ because the normalized divergence is less than the unnormalized divergence. 
\QED
\end{e_rema}

\medskip
{As we have seen in Lemma \ref{lem:entropy_relation_divergence}, we generally have $S_{\rm v}^D(g(\delta) |\vect{X}) \ge S_{\rm v}(\delta |\vect{X}) $ for any $\delta \in [0,1)$ with {$g(\delta) = 2\delta^2 / \ln K$}.
{In particular}, in the case $\delta = 0$, we obtain the following corollary of {Theorems \ref{theo:mean_resolvability} and} \ref{theo:d-VL_resolvability_div}.}
\begin{e_coro}\label{coro:mean_resolvability_relation}
For any general target source ${\vect{X}}$,
\begin{align}
S_{\rm v}^D (0|{\vect{X}}) & = S_{\rm v} ({\vect{X}}). \label{eq:mean_resolvability_formula_divergence}
\end{align}
\QED
\end{e_coro}

Corollary \ref{coro:mean_resolvability_relation} indicates that the $\mathrm{v}_D(0)$-resolvability   $S_{\rm v}^D (0|{\vect{X}})$ coincides with  the $\mathrm{v}$-resolvability $S_{\rm v} ({\vect{X}})$ and is also characterized by the r.h.s.\ of \eqref{eq:mean_resolvability_formula}. 
By \eqref{eq:weak-source-code}, it also implies that $S_{\rm v}^D (0|{\vect{X}}) = R_{\rm v}^*(\vect{X})$, where $R_{\rm v}^*(\vect{X})$ denotes the minimum {error-vanishing} achievable rate by variable-length source codes for $\vect{X}$.

\medskip
\noindent
	\emph{Proof of Theorem \ref{theo:d-VL_resolvability_div}}

\emph{1) Converse Part:}~~~Let $R$ be {$\mathrm{v}_D(\delta)$-achievable}. 
Then, there exists $U^{(L_n)}$ and $\varphi_n$ satisfying {\eqref{eq:d-variable_length_rate_cond_divergence}}  and
\begin{align}
 &\limsup_{n \rightarrow \infty} \delta_n \le \delta, \label{eq:d-variable_length_variational_dist_cond_divergence}   
\end{align}
where we define $\delta_n = D(\ApproxSr||\TargetSr)$ with $\ApproxSr = \varphi_n (U^{(L_n)})$.
Equation \eqref{eq:d-variable_length_variational_dist_cond_divergence} implies that for any given $\gamma >0$, $\delta_n \le \delta + \gamma$ for all {$n \ge n_0$ with some $n_0 >0$}, and therefore
\begin{align}
H_{[\delta + \gamma]}^D(\TargetSr) \le H_{[\delta_n]}^D(\TargetSr)~~~(\forall n \ge n_0) \label{eq:conv_divergence_ineq1}
\end{align}
since $H_{[\delta]}^D(\TargetSr)$ is a nonincreasing function of $\delta$.
Since $P_{\ApproxSr} \in B_{\delta_n}^D(\TargetSr)$, we have
\begin{align}
H_{[\delta_n]}^D(\TargetSr) \le H(\ApproxSr). \label{eq:conv_divergence_ineq2}
\end{align}
On the other hand, it follows from \eqref{eq:entropy-average_length} that
\begin{align}
 H(\ApproxSr) & \le H(U^{(L_n)}) = \E [L_n] +  H(L_n), \label{eq:conv_divergence_ineq3}
\end{align}
where the inequality is due to the fact that $\varphi_n$ is a deterministic mapping and $\ApproxSr = \varphi_n (U^{(L_n)})$.

Combining  \eqref{eq:conv_divergence_ineq1}--\eqref{eq:conv_divergence_ineq3} yields
\begin{align}
H_{[\delta+\gamma]}^D ({\vect{X}}) &= \limsup_{n \rightarrow \infty} \frac{1}{n} H_{[\delta + \gamma]}^D(\TargetSr) \nonumber \\ 
&\le \limsup_{n \rightarrow \infty} \frac{1}{n} \E [L_n]  +  \limsup_{n \rightarrow \infty} \frac{1}{n} H(L_n) \le R, 
\end{align}
where we {used} \eqref{eq:vanishing_length_entropy} and {\eqref{eq:d-variable_length_rate_cond_divergence}} for the last inequality.
Since $\gamma > 0$ is arbitrary, we have 
\begin{align}
\lim_{\gamma \downarrow 0} H_{[\delta+\gamma]}^D ({\vect{X}})  \le R.
\end{align}
\QED

\emph{2) Direct Part:}~~~We modify the achievability scheme in the proof of {the direct part} of Theorem \ref{theo:d-VL_resolvability}. 
Without loss of generality, we assume that $H^* := \lim_{\gamma \downarrow 0} H_{[\delta + \gamma]}^D({\vect{X}})$ is finite ($H^* < + \infty$).
Letting $R = H^* + \mu$, where $\mu > 0$ is an arbitrary constant, we shall show that $R$ is {$\mathrm{v}_D(\delta)$-achievable}.

Let ${V^n}$ be a random variable subject to $P_{V^n} \in B_{\delta + \gamma}^D (\TargetSr)$ satisfying
\begin{align}
H_{[\delta + \gamma]}^D(\TargetSr) + \gamma \ge H(V^n) \label{eq:H_divergence_ineq2}
\end{align}
for {any fixed $\gamma \in (0, \frac{1}{2}]$}.
We can choose $c_n > 0$ {so large} that
\begin{align}
\gamma_0 := \Pr \{ V^n \not\in T_n \} \le \gamma \label{eq:dominant_set_divergence}
\end{align}
where 
\begin{align}
T_n := \left\{ \TargetSeq \in \CdAlphabet^n : \frac{1}{n} \log \frac{1}{P_{V^n}(\TargetSeq)} \le c_n \right\}.
\end{align}
We also define
\begin{align}
\ell (\TargetSeq) :=  \left\lceil \log \frac{1}{P_{V^n}(\TargetSeq)} + n \gamma \right\rceil ~~~\mbox{for}~ \TargetSeq \in T_n.
\end{align}
Letting, for {$m=1, 2, \ldots, \beta_n:=  \lceil n (c_n+\gamma) \rceil$, }
\begin{align}
S_n(m) := \left\{ \TargetSeq \in \CdAlphabet^n : \ell (\TargetSeq) = m \right\},
\end{align}
these sets form a partition of $T_n$:
\begin{align}
\bigcup_{m = 1}^{\beta_n} S_n(m) = T_n. \label{eq:partition2}
\end{align}
We set $L_n$ so that
\begin{align}
\Pr \{ L_n = m\} =  \frac{\Pr\{V^n \in S_n(m)\}}{\Pr\{V^n \in T_n \}} = \frac{\Pr\{V^n \in S_n(m)\}}{1- \gamma_0},
\end{align}
which satisfies
\begin{align}
\sum_{m = 1}^{\beta_n} \Pr \{ L_n = m\} =\frac{\Pr \{ V^n \in T_n \}}{1 - \gamma_0}=1,
\end{align}
and hence the probability distribution of $U^{(L_n)}$ is
\begin{align}
P_{U^{(L_n)}}(\CoinSeq, m) := \Pr \{ U^{(L_n)} = \CoinSeq, L_n = m \} = \frac{\Pr\{V^n \in S_n(m)\}}{(1- \gamma_0)K^m}~~(\forall \CoinSeq \in \SrAlphabet^m).
\end{align}

\medskip
\noindent \emph{Construction of Mapping $\varphi_n: \SrAlphabet^* \rightarrow \CdAlphabet^n$:} \\
%
\quad 
Index the elements in $S_n(m)$ as $\TargetSeq_1, \TargetSeq_2, \ldots,\TargetSeq_{|S_n(m)|}~(m= 1, 2, \ldots, \beta_n )$, where 
\begin{align}
 |S_n(m)| \le K^{m - n\gamma} \label{eq:size_Sn(m)_divergence}
 \end{align}
{(cf.\ \eqref{eq:size_Sn(m)}--\eqref{eq:size_eval})}.
For $i = 1, 2,\ldots, |S_n(m)|$, define $\tilde{A}_i^{(m)} \subset \SrAlphabet^m$ as {the} set of sequences $\CoinSeq \in \SrAlphabet^m$ {so} that
\begin{align}
\sum_{\CoinSeq \in \tilde{A}_i^{(m)}} P_{U^{(L_n)}}(\CoinSeq, m) \le \frac{P_{V^n}(\TargetSeq_i)}{1- \gamma_0} < \sum_{\CoinSeq \in \tilde{A}_i^{(m)}} P_{U^{(L_n)}}(\CoinSeq, m)  + \frac{\Pr\{L_n = m\}}{K^m} \label{eq:prob_set1_divergence}
\end{align}
and
\begin{align}
\tilde{A}_i^{(m)} \cap \tilde{A}_j^{(m)} = \emptyset~~~~(i \neq j). \label{eq:prob_set2_divergence}
\end{align}
If
\begin{align}
\sum_{i =1}^{|S_n(m)|} \sum_{\CoinSeq \in \tilde{A}_i^{(m)}}  P_{U^{(L_n)}}(\CoinSeq, m) <  \frac{1}{1- \gamma_0} \sum_{i =1}^{|S_n(m)|}  P_{V^n}(\TargetSeq_i) = \Pr \{ L_n = m\},
\end{align}
then add {a} $\CoinSeq_i \in \SrAlphabet^m \setminus (\cup_j \tilde{A}_j^{(m)})$ to {obtain}
\begin{align}
A_i^{(m)} = \tilde{A}_i^{(m)} \cup \{ \CoinSeq_i \}
\end{align} 
for $i = 1, 2, \ldots $  in order, 
until it holds that with some $ 1\le c  \le |S_n(m)|$
\begin{align}
\bigcup_{i =1}^{c} A_i^{(m)}\cup \bigcup_{i =c+1}^{|S_n(m)|} \tilde{A}_i^{(m)} = \SrAlphabet^m,
\end{align}
{where $\CoinSeq_1, \CoinSeq_2, \cdots$ are {selected} to be all distinct.}
Since $|\SrAlphabet^m| = K^m$ and
\begin{align}
\sum_{\CoinSeq \in \SrAlphabet^m} P_{U^{(L_n)}}(\CoinSeq, m) =  \sum_{\CoinSeq \in \SrAlphabet^m} \frac{\Pr\{ V^n \in S_n(m)\}}{(1- \gamma_0)K^m} = \Pr\{ L_n = m\},
\end{align}
such $1 \le c \le |S_n(m)|$ always exists.
For simplicity, we set for $i = c+1, {c+2}, \ldots, |S_n(m)|$ 
\begin{align}
A_i^{(m)} = \tilde{A}_i^{(m)}
\end{align}
and for $i = 1, {2}, \ldots, |S_n(m)|$ 
\begin{align}
\varphi_n (\CoinSeq) = \TargetSeq_i ~~\mbox{for}~\CoinSeq \in A_i^{(m)}, 
\end{align}
{which defines the random variable $\tilde{X}^n$ with values in $\mathcal{X}^n$ such that}
\begin{align}
P_{\ApproxSr}(\TargetSeq_i) = \sum_{\CoinSeq \in A_i^{(m)}}  P_{U^{(L_n)}}(\CoinSeq, m)~~~{(\TargetSeq_i \in S_n(m))}, \label{eq:Prob_tildeY_divergence}
\end{align}
{that is, $\tilde{X}^n=\varphi_n(U^{(L_n)})$.}
Notice that by this construction
we have
\begin{align}
 \left|P_{\ApproxSr}(\TargetSeq_i) - \frac{P_{V^n}(\TargetSeq_i)}{1 - \gamma_0} \right| \le \frac{\Pr \{ L_n = m\}}{K^m} = \frac{\Pr \{ V^n \in S_n(m)\}}{(1-\gamma_0)K^m} \label{eq:prob_set3_divergence}
\end{align}
for $i = 1, 2, \ldots, |S_n(m)|; m = 1, 2, \ldots, \beta_n$, and
\begin{align}
\Pr \{ \ApproxSr \not\in T_n \} = 0 ~~~ \mbox{and} ~~~\Pr \{ V^n \not\in T_n \} \le \gamma. \label{eq:Prob_tildeY_divergence2}
\end{align}

\medskip
\noindent \emph{Evaluation of Average Length:} \\
\quad
The average length $\E [L_n]$ is evaluated as follows:
\begin{align}
\E[L_n] &= \sum_{m=1}^{\beta_n}  \sum_{\CoinSeq \in \SrAlphabet^m} P_{U^{(L_n)}}(\CoinSeq, m) \cdot m \nonumber \\
&=\sum_{m=1}^{\beta_n} \sum_{i=1}^{|S_n(m)|} \sum_{\CoinSeq \in A_i^{(m)}} P_{U^{(L_n)}}(\CoinSeq, m) \cdot m \nonumber \\
&= \sum_{m=1}^{\beta_n} \sum_{\TargetSeq_i \in S_n(m)} P_{\ApproxSr} (\TargetSeq_i) \cdot m, \label{eq:average_length_divergence}
\end{align}
where we have used $\SrAlphabet^m = \bigcup_{i=1}^{|S_n(m)|} A_i^{(m)}$ and \eqref{eq:Prob_tildeY_divergence}.
For $\TargetSeq_i \in S_n(m)$ we obtain from {\eqref{eq:prob_set3_divergence} and the right inequality of \eqref{eq:Prob_tildeY_divergence2}}
\begin{align}
P_{\ApproxSr}(\TargetSeq_i) &\le \frac{ P_{V^n}(\TargetSeq_i)}{1- \gamma_0} + \frac{\Pr\{ V^n \in S_n(m)\}}{(1-\gamma_0)K^m}  \nonumber \\
 & = \left(1+ \frac{\gamma_0}{1- \gamma_0} \right) \left( P_{V^n}(\TargetSeq_i) + \frac{\Pr\{V^n \in S_n(m) \}}{K^m} \right)\nonumber \\
 & \le  \left(1+ \frac{\gamma_0}{1- \gamma_0} \right) \left( 1 + \frac{1}{P_{V^n}(\TargetSeq_i)K^m} \right) P_{V^n}(\TargetSeq_i)  \nonumber \\
 & \le \left(1+ 2 \gamma \right) \left( 1 + \frac{1}{K^{n\gamma}} \right) P_{V^n}(\TargetSeq_i) , \label{eq:bound_P_tildeY_divergence}
\end{align}
where, to derive the last inequality, we have used the fact ${0 \le} \, \gamma_0 \le \gamma \le \frac{1}{2}$ {and
\begin{align}
P_{V^n}(\TargetSeq_i) \ge K^{-(m - n\gamma)}~~~(\forall \TargetSeq_i \in S_n(m)).
\end{align}
It should be noticed that \eqref{eq:bound_P_tildeY_divergence} also implies that 
\begin{align}
P_{\ApproxSr}(\TargetSeq) \le \left(1+ 2 \gamma \right) \left( 1 + \frac{1}{K^{n\gamma}} \right) P_{V^n}(\TargetSeq)~~~(\forall \TargetSeq \in \mathcal{X}^n) \label{eq:bound_P_tildeY_divergence2}
\end{align}
since $P_{\ApproxSr}(\TargetSeq) = 0 $ for $\TargetSeq \not\in T_n = \bigcup_{m=1}^{\beta_n} S_n (m)$.}
Plugging the inequality
\begin{align}
m \le \log \frac{1}{P_{V^n}(\TargetSeq_i)} + n\gamma + 1 ~~(\forall \TargetSeq_i \in S_n(m))
\end{align}
and \eqref{eq:bound_P_tildeY_divergence} into \eqref{eq:average_length_divergence}, we obtain
\begin{align}
\E[L_n]  &\le (1+ 2 \gamma) \left( 1 + \frac{1}{K^{n\gamma}} \right) \sum_{m=1}^{\beta_n} \sum_{\TargetSeq_i \in S_n(m)} P_{V^n}(\TargetSeq_i) \left(\log \frac{1}{P_{V^n}(\TargetSeq_i)} + n\gamma + 1\right) \nonumber \\
&\le (1+ 2 \gamma)\left( 1 + \frac{1}{K^{n\gamma}} \right) \left( H(V^n) + n\gamma + 1 \right). \label{eq:average_length_divergence2}
\end{align}
Thus, we obtain from \eqref{eq:average_length_divergence2}
\begin{align}
\limsup_{n \rightarrow \infty} \frac{1}{n} \E [L_n] &\le (1+ 2 \gamma)  \left\{ \limsup_{n \rightarrow \infty} \frac{1}{n} H(V^n) + \gamma \right\} \nonumber \\
&\le   (1+ 2 \gamma) \left\{ \limsup_{n \rightarrow \infty} \frac{1}{n} H_{[\delta + \gamma]}^D (\TargetSr) + 2 \gamma \right\} \nonumber \\
&\le  (1+ 2 \gamma) (  H^*  + 2 \gamma), \label{eq:rate_cond_satisfy_divergence}
\end{align}
where the second inequality follows from \eqref{eq:H_divergence_ineq2}.
{Since we have assume that $H^*$ is finite and $\gamma \in (0, \frac{1}{2}]$ is arbitrary, the r.h.s.\ of \eqref{eq:rate_cond_satisfy_divergence} can be made as close to $H^*$ as desired.
Therefore, for all sufficiently small $\gamma > 0$ we obtain
\begin{align}
\limsup_{n \rightarrow \infty} \frac{1}{n} \E [L_n] &\le H^*  + \mu = R \label{eq:rate_cond_satisfy_divergence2}
\end{align}}

\medskip
\noindent \emph{Evaluation of Divergence:} \\
\quad 
The divergence $D(\ApproxSr || \TargetSr)$ can be rewritten as
\begin{align}
D(\ApproxSr || \TargetSr) = D(\ApproxSr || V^n) + \E \left[ \log \frac{P_{V^n}(\ApproxSr)}{P_{\TargetSr}(\ApproxSr)}\right]. \label{eq:divergence2}
\end{align}
In view of \eqref{eq:bound_P_tildeY_divergence}, we obtain
\begin{align}
 D(\ApproxSr || V^n) &= \sum_{m = 1}^{\beta_n} \sum_{\TargetSeq \in S_n(m)} P_{\ApproxSr}(\TargetSeq) \log \frac{P_{\ApproxSr}(\TargetSeq)}{P_{V^n}(\TargetSeq)} \nonumber \\
 &\le \sum_{m = 1}^{\beta_n} \sum_{\TargetSeq \in S_n(m)} P_{\ApproxSr}(\TargetSeq) \log \left(1+ 2 \gamma \right) \left( 1 + \frac{1}{K^{n\gamma}} \right) 
 \nonumber \\
  &\le  \frac{2 \gamma}{\ln K}  \, {+ \log  \left( 1 + \frac{1}{K^{n\gamma}} \right)} \label{eq:divergence_UB1}
\end{align}
and 
\begin{align}
 \E \left[ \log \frac{P_{V^n}(\ApproxSr)}{P_{\TargetSr}(\ApproxSr)}\right] 
 &= \sum_{\TargetSeq \in \mathcal{X}^n} P_{\ApproxSr}(\TargetSeq) \log  \frac{P_{V^n}(\TargetSeq)}{P_{\TargetSr}(\TargetSeq)} \nonumber \\
 &\le \left(1+ 2 \gamma \right) \left( 1 + \frac{1}{K^{n\gamma}} \right) D(V^n || \TargetSr) \nonumber \\
 &\le \left(1+ 2 \gamma \right)  (\delta + \gamma) \left( 1 + \frac{1}{K^{n\gamma}} \right), \label{eq:divergence_UB2}
\end{align}
where to obtain the last inequality we used the fact that $P_{V^n} \in B_{\delta+\gamma}^D (\TargetSr)$.
Plugging \eqref{eq:divergence_UB1} and \eqref{eq:divergence_UB2} into \eqref{eq:divergence2} yields 
\begin{align}
\limsup_{n \rightarrow \infty} D(\ApproxSr || \TargetSr) &\le \frac{2 \gamma}{\ln K} + (1+ 2 \gamma )(\delta + \gamma) \nonumber \\
 &\le \delta + { \gamma (2 \delta + 5)} , \label{eq:divergence5}
\end{align}
where we have used the fact that $\frac{2 \gamma}{\ln K} \le 3 \gamma$ for all $K \ge 2$ {and the assumption $0 < \gamma \le \frac{1}{2}$} to derive the last inequality.
Since $\gamma \in (0, \frac{1}{2}]$ is arbitrary and we have \eqref{eq:rate_cond_satisfy_divergence2}, $R$ is {$\mathrm{v}_D(\delta)$-achievable}.
\QED

\medskip\section{Mean and Variable-Length Channel Resolvability} \label{sec:channel_resolvability}

{So far we have studied the problem of  \emph{source resolvability}, whereas} the problem of \emph{channel resolvability} has been introduced by Han and Verd\'u \cite{Han-Verdu93} to investigate the capacity of identification codes \cite{Ahlswede-Dueck1989}.
In the conventional problem {of this kind}, a target \emph{output distribution} $P_{Y^n}$ via a channel $W^n$ {due to an input $X^n$} is approximated by encoding the fixed-length uniform random number ${U_{M_n}}$ as a channel input.  
In this section, we generalize the problem of {such} channel resolvability to {in} the variable-length setting.

\subsection{Definitions}

Let $\mathcal{X}$ and $\mathcal{Y}$ be finite or countably infinite alphabets.
Let $\vect{W} = \{W^n \}_{n=1}^\infty$ be a general  channel, where $W^n : \mathcal{X}^n \rightarrow \mathcal{Y}^n$ denotes a stochastic mapping.
We denote by $\vect{Y} = \{ Y^n \}_{n=1}^\infty$ the output process from $\vect{W}$ due to the input process $\vect{X}= \{ X^n \}_{n=1}^\infty$, {where $X^n$ and  $Y^n$ take values in $\mathcal{X}^n$ and $\mathcal{Y}^n$, respectively}. 
Again, we do not impose any assumptions such as stationarity or ergodicity on either $\vect{X}$ or $\vect{W}$.
As in the previous sections, we will identify $X^n$ and $Y^n$ with their probability distributions $P_{X^n}$ and $P_{Y^n}$, respectively, and these symbols are used interchangeably.

In this section, we consider several {types of} problems of approximating a target output distribution $P_{Y^n}$.
The first one is the problem of \emph{mean}-resolvability \cite{Han-Verdu93}, in which the channel input is allowed to be an arbitrary general source.
\medskip
\begin{e_defin}[{$\delta$-Mean} Channel Resolvability: Variational Distance] \label{def:mean_ch_achievable_rate}
Let $\delta \in [0, 1)$ be fixed arbitrarily.
A resolution rate $R \ge 0$ is said to be \emph{$\delta$-{mean achievable} for $\vect{X}$} {(under the variational distance)} if there exists a general source $\tilde{\vect{X}} = \{ \tilde{X}^n\}_{n=1}^\infty$ satisfying
\begin{align}
 \limsup_{n \rightarrow \infty} \frac{1}{n} H(\tilde{X}^n) &\le R, \label{eq:mean_rate_cond}  \\ 
 \limsup_{n \rightarrow \infty} d(P_{Y^n}, P_{\tilde{Y}^n}) &\le \delta, \label{eq:mean_variational_dist_cond}   
\end{align}
where $\tilde{Y}^n $ denotes the output via $W^n$ due to the input $\tilde{X}^n$.
The infimum of all $\delta$-mean achievable rates for $\vect{X}$:
\begin{align}
\overline{S}_{\rm v} (\delta | \vect{X}, \vect{W}):= \inf \{ R : ~ R ~\mbox{is $\delta$-mean achievable for}~ \vect{X}\} \label{eq:mean_opt_rate}
\end{align}
is {referred to as} the \emph{$\delta$-mean resolvability} for $\vect{X}$.
We also define the \emph{$\delta$-mean resolvability for the worst input} as
\begin{align}
\overline{S}_{\rm v} (\delta | \vect{W}):= \sup_{\vect{X}}  \overline{S}_{\rm v} (\delta | \vect{X}, \vect{W}). \label{eq:mean_capacity}
\end{align}
\QED
\end{e_defin}

\medskip
{On the other hand, we may also consider the problem of variable-length channel resolvability.}
Here, the variable-length uniform random number $U^{(L_n)}$ is defined as in the foregoing sections.
Consider the problem of approximating the target output distribution $P_{Y^n}$ {via} $W^n$ {due to $X_n$} by using another input $\tilde{X}^n=\varphi_n(U^{(L_n)})$ with a deterministic mapping $\varphi_n :  \mathcal{U}^*  \rightarrow \mathcal{X}^n$.
\begin{e_defin}[{$\delta$-Variable-Length} Channel Resolvability: Variational Distance] \label{def:VL_achievable_rate}
Let $\delta \in [0, 1)$ be fixed arbitrarily.
A resolution rate $R \ge 0$ is said to be \emph{$\delta$-variable-length achievable} {or simply $\mathrm{v}(\delta)$-achievable} for $\vect{X}$ {(under the variational distance)} if there exists a variable-length uniform random number $U^{(L_n)}$ and a deterministic mapping $\varphi_n : \mathcal{U}^* \rightarrow \mathcal{X}^n$ satisfying
\begin{align}
 \limsup_{n \rightarrow \infty} \frac{1}{n} \E [L_n] &\le R, \label{eq:VL_ch_rate_cond}  \\ 
 \limsup_{n \rightarrow \infty} d(P_{Y^n}, P_{\tilde{Y}^n}) &\le \delta, \label{eq:VL_ch_variational_dist_cond}   
\end{align}
where $\E[\cdot]$ denotes the expected value and $\tilde{Y}^n $ denotes the output via $W^n$ due to the input $\tilde{X}^n = \varphi_n(U^{(L_n)})$.
The infimum of all {$\mathrm{v}(\delta)$-achievable} rates for $\vect{X}$:
\begin{align}
S_{\rm v} (\delta | \vect{X}, \vect{W}):= \inf \{ R : ~ R ~\mbox{is {$\mathrm{v}(\delta)$-achievable} for}~ \vect{X}\} \label{eq:VL_ch_opt_rate}
\end{align}
is called the \emph{$\delta$-variable-length channel resolvability} {or simply $\mathrm{v}(\delta)$-channel resolvability} for $\vect{X}$.
We also define the \emph{$\delta$-variable-length channel resolvability {or simply {$\mathrm{v}(\delta)$-channel resolvability}} for the worst input} as
\begin{align}
S_{\rm v} (\delta | \vect{W}):= \sup_{\vect{X}}  S_{\rm v} (\delta | \vect{X}, \vect{W}). \label{eq:variable_length_capacity}
\end{align}
\QED
\end{e_defin}

When $W^n$ is the \emph{identity mapping}, the problem of channel resolvability reduces to that of source resolvability, which had been investigated in the foregoing sections. 
In this sense, the problem of channel resolvability is a generalization of the problem of source resolvability. 

\medskip
Similarly to the problem of source resolvability, we may also use the divergence between the target output distribution $P_{Y^n}$ and the approximated output distribution $P_{\tilde{Y}^n}$ as {the approximation measure}.
\medskip
\begin{e_defin}[{$\delta$-Mean} Channel Resolvability: Divergence] \label{def:VL_achievable_rate_div}
Let $\delta \ge 0$ be fixed arbitrarily.
A resolution rate $R \ge 0$ is said to be \emph{$\delta$-{mean achievable} for $\vect{X}$} {(under the divergence)} if there exists a general source $\tilde{\vect{X}} = \{ \tilde{X}^n\}_{n=1}^\infty$ satisfying
\begin{align}
 \limsup_{n \rightarrow \infty} \frac{1}{n} H(\tilde{X}^n) &\le R, \label{eq:mean_rate_cond_div}  \\ 
 \limsup_{n \rightarrow \infty} D(\tilde{Y}^n || Y^n ) &\le \delta, \label{eq:mean_variational_dist_con_div}   
\end{align}
where $\tilde{Y}^n $ denotes the output via $W^n$ due to the input $\tilde{X}^n$.
The infimum of all $\delta$-mean achievable rates for $\vect{X}$:
\begin{align}
\overline{S}_{\rm v}^D (\delta | \vect{X}, \vect{W}):= \inf \{ R : ~ R ~\mbox{is $\delta$-mean achievable for}~ \vect{X}\} \label{eq:mean_opt_rate_div}
\end{align}
is {referred to as} the \emph{$\delta$-mean channel resolvability} for $\vect{X}$.
We also define the \emph{$\delta$-mean channel resolvability for the worst input} as
\begin{align}
\overline{S}_{\rm v}^D (\delta | \vect{W}):= \sup_{\vect{X}}  \overline{S}_{\rm v}^D (\delta | \vect{X}, \vect{W}). \label{eq:mean_capacity_div}
\end{align}
\QED
\end{e_defin}

\medskip
\begin{e_defin}[{$\delta$-Variable-Length} Channel Resolvability: Divergence] \label{def:mean_ch_achievable_rate_div}
Let $\delta \ge 0$ be fixed arbitrarily.
A resolution rate $R \ge 0$ is said to be \emph{$\delta$-variable-length achievable} {or simply $\mathrm{v}_D (\delta)$-achievable} for $\vect{X}$ {(under the divergence)} if there exists a variable-length uniform random number $U^{(L_n)}$ and a deterministic mapping $\varphi_n : \mathcal{U}^* \rightarrow \mathcal{X}^n$ satisfying
\begin{align}
 \limsup_{n \rightarrow \infty} \frac{1}{n} \E [L_n] &\le R, \label{eq:VL_ch_rate_cond_div}  \\ 
 \limsup_{n \rightarrow \infty} D(\tilde{Y}^n || Y^n ) &\le \delta, \label{eq:VL_cond_div}   
\end{align}
where $\E[\cdot]$ denotes the expected value and $\tilde{Y}^n $ denotes the output via $W^n$ due to the input $\tilde{X}^n = \varphi_n(U^{(L_n)})$.
The infimum of all {$\mathrm{v}_D(\delta)$-achievable} rates for $\vect{X}$:
\begin{align}
S_{\rm v}^D (\delta | \vect{X}, \vect{W}):= \inf \{ R : ~ R ~\mbox{is {$\mathrm{v}_D (\delta)$-achievable} for}~ \vect{X}\} \label{eq:VL_ch_opt_rate_div}
\end{align}
is called the \emph{$\delta$-variable-length channel resolvability} {or simply $\mathrm{v}_D(\delta)$-channel resolvability} for $\vect{X}$.
We also define the \emph{$\delta$-variable-length channel resolvability {or simply $\mathrm{v}_D(\delta)$-channel resolvability} for the worst input} as
\begin{align}
S_{\rm v}^D (\delta | \vect{W}):= \sup_{\vect{X}}  S_{\rm v}^D (\delta | \vect{X}, \vect{W}). \label{eq:variable_length_capacity_div}
\end{align}
\QED
\end{e_defin}

\medskip
\begin{e_rema} \label{rema:mean-VL-relation}
Since the outputs of a deterministic mapping $\tilde{X}^n = \varphi_n(U^{(L_n)})$ form a general source $\tilde{\vect{X}}$,  it holds that
\begin{align}
\overline{S}_{\rm v} (\delta | \vect{X}, \vect{W}) &\le S_{\rm v} (\delta | \vect{X}, \vect{W})~~~(\delta \in [0,1)), \label{eq:mean_VL_general_relation}\\
\overline{S}_{\rm v}^D (\delta | \vect{X}, \vect{W}) &\le S_{\rm v}^D (\delta | \vect{X}, \vect{W})~~(\delta \ge 0) \label{eq:mean_VL_general_relation_div}
 \end{align} 
 for any general source $\vect{X}$ and general channel $\vect{W}$. 
 These relations lead to the analogous relation for the $\delta$-mean/variable-length channel resolvability for the worst input:
 \begin{align}
\overline{S}_{\rm v} (\delta |\vect{W}) &\le S_{\rm v} (\delta | \vect{W})~~~(\delta \in [0,1)), \\
\overline{S}_{\rm v}^D (\delta |\vect{W}) &\le S_{\rm v}^D (\delta | \vect{W})~~(\delta \ge 0). 
 \end{align} 
 \QED
\end{e_rema}

\subsection{General Formulas}

For a given general source $\vect{X} = \{ X^n \}_{n=1}^\infty $ and a general channel $\vect{W} = \{W^n\}_{n=1}^\infty $, let $\vect{Y} = \{ Y^n \}_{n=1}^\infty$ be the channel output via $\vect{W}$ due to input $\vect{X}$. 
We define
\begin{align}
H_{[\delta], W^n}(X^n) = \inf_{P_{V^n} \in B_{\delta} (X^n, W^n)} H(V^n), \\
H_{[\delta], W^n}^D(X^n) = \inf_{P_{V^n} \in B_{\delta}^D (X^n, W^n)} H(V^n)
\end{align}
where $H(V^n)$ denotes the Shannon entropy of $V^n$
and $ B_\delta (X^n, W^n)$ and $ B_\delta^D (X^n, W^n)$ are defined as 
\begin{align}
B_\delta (X^n, W^n) = \left\{ P_{V^n} \in \mathcal{P}(\mathcal{X}^n): d(P_{Y^n}, P_{Z^n}) \le \delta \right\} \label{eq:set_B_delta}
\end{align}
and 
\begin{align}
B_\delta^D (X^n, W^n) = \left\{ P_{V^n} \in \mathcal{P}(\mathcal{X}^n): D(Z^n||Y^n) \le \delta \right\}, \label{eq:set_B_delta_div}
\end{align}
respectively, with $Z^n$ defined as the output from $W^n$ due to the input $V^n$.
Both $H_{[\delta], W^n}(X^n) $ and $H_{[\delta], W^n}^D (X^n) $ are nonincreasing functions of $\delta$.
In addition, we define
\begin{align}
H_{[\delta], \vect{W}}(\vect{X}) &= \limsup_{n \rightarrow \infty} \frac{1}{n} H_{[\delta], W^n}(X^n), \\
H_{[\delta], \vect{W}}^D (\vect{X}) &= \limsup_{n \rightarrow \infty} \frac{1}{n} H_{[\delta], W^n}^D (X^n),
\end{align}
which play an important role in characterizing the $\delta$-mean/variable-length channel resolvability.

We show the general formulas for the $\delta$-mean/variable-length channel resolvability.
\begin{e_theo}[With Variational Distance]  \label{theo:d-VL_ch_resolvability}
For any input process $\vect{X}$ and any general channel $\vect{W}$, 
\begin{align}
\overline{S}_{\rm v} (\delta | \vect{X}, \vect{W}) = S_{\rm v} (\delta | \vect{X}, \vect{W}) = \lim_{\gamma \downarrow 0} H_{[\delta+\gamma], \vect{W}}(\vect{X})~~~(\delta \in [0,1)). 
 \label{eq:d-VL_ch_formula}
\end{align}
In particular,
\begin{align}
\overline{S}_{\rm v} (\delta | \vect{W}) = S_{\rm v} (\delta | \vect{W}) &= \sup_{\vect{X}} \lim_{\gamma \downarrow 0} H_{[\delta+\gamma], \vect{W}}(\vect{X}) ~~~(\delta \in [0,1)). 
\label{eq:d-VL_ch_formula2}
\end{align}
\QED
\end{e_theo}

\begin{e_theo}[With Divergence]  \label{theo:d-VL_ch_resolvability_div}
For any input process $\vect{X}$ and any general channel $\vect{W}$, 
\begin{align}
\overline{S}_{\rm v}^D (\delta | \vect{X}, \vect{W}) = S_{\rm v}^D (\delta | \vect{X}, \vect{W}) = \lim_{\gamma \downarrow 0} H_{[\delta+\gamma], \vect{W}}^D(\vect{X})~~~(\delta \ge 0). 
 \label{eq:d-VL_ch_formula_div}
\end{align}
In particular,
\begin{align}
\overline{S}_{\rm v}^D (\delta | \vect{W})  = S_{\rm v}^D (\delta | \vect{W}) = \sup_{\vect{X}} \lim_{\gamma \downarrow 0} H_{[\delta+\gamma], \vect{W}}^D (\vect{X}) ~~~(\delta \ge 0). 
\label{eq:d-VL_ch_formula2_div}
\end{align}
\QED
\end{e_theo}

\begin{e_rema}
It can be easily verified that the variational distance satisfies 
\begin{align}
d(P_{Y^n}, P_{Z^n}) &\le d(P_{X^n}, P_{V^n}), 
\end{align}
and therefore we have $B_{\delta}(X^n) \subseteq B_{\delta}(X^n, W^n)$.
This relation and formulas \eqref{eq:d-mean_resolvability_formula} and \eqref{eq:d-VL_ch_formula} indicate that 
\begin{align}
S_{\rm v} (\delta | \vect{X}, \vect{W}) \le S_{\rm v} (\delta | \vect{X})~~~(\delta \in[0,1))
\end{align}
for any given channel $\vect{W}$.
Likewise, it is well-known that the divergence satisfies the \emph{data processing inequality} $D(\tilde{Y}^n || Y^n)\le D(\tilde{X}^n || X^n)$ \cite{Csiszar-Korner2011}, and formulas \eqref{eq:d-VL_resolvability_formula_div} and \eqref{eq:d-VL_ch_formula_div} lead to 
\begin{align}
S_{\rm v}^D (\delta | \vect{X}, \vect{W}) \le S_{\rm v}^D (\delta | \vect{X})~~~(\delta \ge 0)
\end{align}
regardless of a channel $\vect{W}$.
\QED
\end{e_rema}

{
\begin{e_rema} \label{rema:source_resolvability_recapture}
It is obvious that Theorems \ref{theo:d-VL_ch_resolvability} and \ref{theo:d-VL_ch_resolvability_div} reduce to Theorems \ref{theo:d-VL_resolvability} and \ref{theo:d-VL_resolvability_div}, respectively, when the channel $\vect{W}$ is the identity mapping.
Precisely, for the identity mapping $\vect{W}$, the $\delta$-mean resolvability and the  $\mathrm{v}(\delta)$-channel resolvability for $\vect{X}$ are given by
\begin{align}
\overline{S}_{\rm v} (\delta | \vect{X}) = S_{\rm v} (\delta | \vect{X}) = \lim_{\gamma \downarrow 0} H_{[\delta+ \gamma]}(\vect{X}), \label{eq:source_resolvability_recapture}
\end{align}
where $\overline{S}_{\rm v} (\delta | \vect{X})$ denotes the $\delta$-mean resolvability $\overline{S}_{\rm v} (\delta | \vect{X}, \vect{W}) $ for the identity mapping $\vect{W}$.
The analogous relationship holds under the divergence:
\begin{align}
\overline{S}_{\rm v}^D (\delta | \vect{X}) = S_{\rm v}^D (\delta | \vect{X}) = \lim_{\gamma \downarrow 0} H_{[\delta+ \gamma]}^D(\vect{X}), \label{eq:source_resolvability_recapture2}
\end{align}
where $\overline{S}_{\rm v}^D (\delta | \vect{X})$ denotes the $\delta$-mean resolvability $\overline{S}_{\rm v}^D (\delta | \vect{X}, \vect{W}) $ for the identity mapping $\vect{W}$.
\QED
\end{e_rema}}

\medskip
\noindent
\emph{Proof of Theorems \ref{theo:d-VL_ch_resolvability} and \ref{theo:d-VL_ch_resolvability_div}}

\emph{1) Converse Part:}~~~Because of the general relationship \eqref{eq:mean_VL_general_relation}, to prove the converse part of Theorem \ref{theo:d-VL_ch_resolvability}, it suffices to show that
\begin{align}
\overline{S}_{\rm v} (\delta | \vect{X}, \vect{W}) &\ge \lim_{\gamma \downarrow 0} H_{[\delta+\gamma], \vect{W}}(\vect{X}). \label{eq:d-VL_ch_formula_LB} 
\end{align}
Let $R$ be $\delta$-mean achievable for $\vect{X}$ under the variational distance. Then, there exists a general source $\tilde{\vect{X}}$ satisfying \eqref{eq:mean_rate_cond} and
\begin{align}
 \limsup_{n \rightarrow \infty} \delta_n \le \delta, \label{eq:mean_variational_dist_con2}
 \end{align}
where $\delta_n := d(P_{Y^n}, P_{\tilde{Y}^n})$.  
Fixing $\gamma > 0$ arbitrarily, we have $\delta_n \le \delta + \gamma$ for all $n \ge n_0$ with some $n_0 > 0$ and then
\begin{align}
H_{[\delta+\gamma], W^n}(X^n) \le H_{[\delta_n], W^n}(X^n)~~~(n \ge n_0)
\end{align}
since $H_{[\delta], W^n}(\TargetSr)$ is a nonincreasing function of $\delta$.
Since $P_{\tilde{X}^n} \in B_{\delta_n}(X^n, W^n)$, we have $H_{[\delta_n], W^n} (X^n) \le H(\tilde{X}^n)$.
Thus, we obtain from \eqref{eq:mean_rate_cond} 
\begin{align}
H_{[\delta+\gamma], \vect{W}}(\vect{X}) \le \limsup_{n \rightarrow \infty} \frac{1}{n} H (\tilde{X}^n) \le R.
\end{align}
Since $\gamma >0$ is an arbitrary constant, this implies that we have \eqref{eq:d-VL_ch_formula_LB}.

 The converse part of Theorem \ref{theo:d-VL_ch_resolvability_div} can be proven in an analogous way with due modifications.
 
 \medskip
 \emph{2) Direct Part:}~~~Because of the general relationship \eqref{eq:mean_VL_general_relation}, to prove the direct part (achievability) of Theorem \ref{theo:d-VL_ch_resolvability}, it suffices to show that for any fixed $\gamma > 0$ the {resolution} rate
\begin{align}
R =  \lim_{\gamma \downarrow 0} H_{[\delta+\gamma], \vect{W}}(\vect{X}) + 3 \gamma \label{eq:d-VL_ch_formula_UB} 
\end{align}
is {$\mathrm{v}(\delta)$-achievable} for $\vect{X}$ under the variational distance. 

Let $P_{V^n} \in B_{\delta + \gamma} (X^n, W^n)$ be a source satisfying 
\begin{align}
H (V^n) \le  H_{[\delta+\gamma], W^n} (X^n) + \gamma. \label{eq:HV_setting}
\end{align}
Then, by the same argument to derive \eqref{eq:rate_cond_satisfy} and \eqref{eq:distance_UB1} developed in the proof of the direct part of Theorem \ref{theo:d-VL_resolvability}, we can construct {a variable-length uniform random number $U^{(L_n)}$ and} a {deterministic} mapping $\varphi_n : \mathcal{U}^* \rightarrow \mathcal{X}^n$ satisfying 
\begin{align}
\limsup_{n \rightarrow \infty} \frac{1}{n} \E[L_n] \le  \lim_{\gamma \downarrow 0} H_{[\delta+\gamma], \vect{W}}(\vect{X}) + 3 \gamma = R
\end{align}
and 
\begin{align}
d(P_{\tilde{X}^n}, P_{V^n}) \le \frac{1}{2} K^{- n\gamma} + \gamma, \label{eq:distance_UB1b}
\end{align}
{where $\tilde{X}^n = \varphi_n(U^{(L_n)})$}.
Let $Z^n$ denote the output random variable from $W^n$ due to the input $V^n$.
Then, {letting  $\tilde{Y}^n$ be the channel output via channel $W^n$ due to input $\tilde{X}^n$},  we can evaluate  
$d(P_{\tilde{Y}^n}, P_{Z^n})$ as
\begin{align}
d(P_{\tilde{Y}^n}, P_{Z^n}) &= \frac{1}{2} \sum_{\vect{y} \in \mathcal{Y}^n} |P_{\tilde{Y}^n}(\vect{y})- P_{Z^n}(\vect{y})| \nonumber \\
& = \frac{1}{2} \sum_{\vect{y} \in \mathcal{Y}^n}  \left| \sum_{\vect{x} \in \mathcal{X}^n} W(\vect{y} | \vect{x}) \big(P_{\tilde{X}^n}(\vect{x})- P_{V^n}(\vect{x}) \big) \right| \nonumber \\
&  \le  \frac{1}{2} \sum_{\vect{y} \in \mathcal{Y}^n} \sum_{\vect{x} \in \mathcal{X}^n} W(\vect{y} | \vect{x}) \left|  P_{\tilde{X}^n}(\vect{x})- P_{V^n}(\vect{x}) \right| \nonumber \\
& = d(P_{\tilde{X}^n}, P_{V^n}) \le \frac{1}{2}  K^{-n\gamma} + \gamma. \label{eq:variational_dist_UB1}
\end{align}
Thus, we obtain
\begin{align}
\limsup_{n \rightarrow \infty} d(P_{Y^n}, P_{\tilde{Y}^n})  &\le\limsup_{n \rightarrow \infty} d(P_{Y^n}, P_{Z^n}) +  \limsup_{n \rightarrow \infty} d(P_{\tilde{Y}^n}, P_{Z^n}) \nonumber \\
&\le \delta + 2 \gamma,
\end{align}
where we have used the fact $P_{V^n} \in B_{\delta+\gamma} (X^n, W^n)$ to derive the last inequality. 
Since $\gamma > 0$ is an arbitrary constant, we can conclude that $R$ is {$\mathrm{v}(\delta)$-achievable} for $\vect{X}$.  

  The direct part of Theorem \ref{theo:d-VL_ch_resolvability_div} can be proven in the same way as Theorem \ref{theo:d-VL_resolvability_div} with due modifications.
  Fixing $P_{V^n} \in B_{\delta + \gamma}^D(X^n, W^n)$ and using the encoding scheme {developed} in the proof of Theorem \ref{theo:d-VL_resolvability_div}, the evaluation of the average length rate is exactly the same and we {can} obtain \eqref{eq:rate_cond_satisfy_divergence2}.
  A key step is to evaluate the divergence $D(\tilde{Y}^n || Y^n)$, which can be rewritten as
 \begin{align}
D(\tilde{Y}^n || Y^n) = D(\tilde{Y}^n || Z^n) + \E \left[ \log \frac{P_{Z^n}(\tilde{Y}^n)}{P_{Y^n}(\tilde{Y}^n)}\right]. \label{eq:output_divergence}
\end{align}
The first term on the r.h.s.\ can be bounded as
\begin{align}
 D(\tilde{Y}^n || Z^n) &\le D(\tilde{X}^n || V^n)
  \le  \frac{2 \gamma}{\ln K}  + \log  \left( 1 + \frac{1}{K^{n\gamma}} \right)
  \end{align}
  as in \eqref{eq:divergence_UB1}, where the left inequality is due to the data processing inequality.
  Similarly to the derivation of \eqref{eq:divergence_UB2}, the second term can be bounded as
\begin{align}
 \E \left[ \log \frac{P_{Z^n}(\tilde{Y}^n)}{P_{Y^n}(\tilde{Y}^n)}\right] 
&= \sum_{\vect{y} \in \mathcal{Y}^n} \sum_{\vect{x} \in \mathcal{X}^n} P_{\tilde{X}^n}(\vect{x}) W^n(\vect{y}|\vect{x})\log  \frac{P_{Z^n}(\vect{y})}{P_{Y^n}(\vect{y})} \nonumber \\
  &\le \left(1+ 2 \gamma \right) \left( 1 + \frac{1}{K^{n\gamma}} \right) \sum_{\vect{y} \in \mathcal{Y}^n}  \sum_{\vect{x} \in \mathcal{X}^n} P_{V^n}(\vect{x}) W^n(\vect{y}|\vect{x})\log  \frac{P_{Z^n}(\vect{y})}{P_{Y^n}(\vect{y})}  \nonumber \\
 &= \left(1+ 2 \gamma \right) \left( 1 + \frac{1}{K^{n\gamma}} \right)  D(Z^n||Y^n),  \label{eq:ch_divergence_UB2}
\end{align}
where {we have used {\eqref{eq:bound_P_tildeY_divergence2}}. Here,} $D(Z^n||Y^n) \le \delta + \gamma$ because $Z^n$ is the output via $W^n$ due to the input $V^n \in B_{\delta + \gamma}^D(X^n, W^n)$. 
The rest of the steps is the same as in the proof of Theorem \ref{theo:d-VL_resolvability_div}.
 \QED
 
\medskip\section{Second-Order {Variable-Length} Channel Resolvability} \label{sec:second-order}

\subsection{Definitions}

We {now} turn to considering the \emph{second-order} resolution rates \cite{Nomura-Han2013,Watanabe-Hayashi2014}. First, we consider the variable-length resolvability based on the variational distance.
 
\begin{e_defin}[$(\delta, R)$-{Variable-Length} Channel Resolvability: Variational Distance]
A second-order resolution rate {$L \in (-\infty, + \infty)$} is said to be \emph{$(\delta, R)$-{variable-length achievable}} {(under {the} variational distance)} for $\vect{X}$ {with $\delta \in [0, 1)$} if there exists a variable-length uniform random number $U^{(L_n)}$ and
a deterministic mapping $\varphi_n : \SrAlphabet^* \rightarrow \CdAlphabet^n$ satisfying
\begin{align}
 \limsup_{n \rightarrow \infty} \frac{1}{\sqrt{n}} \left( \E [L_n] - nR \right) &\le L, \label{eq:dR-variable_length_rate_cond}  \\ 
 \limsup_{n \rightarrow \infty} d(P_{Y^n}, P_{\tilde{Y}^n}) &\le \delta, \label{eq:dR-variable_length_variational_dist_cond}   
\end{align}
where $\tilde{Y}^n$ denotes the output via $W^n$ due to the input  $\ApproxSr = \varphi_n(U^{(L_n)})$.
The infimum of all $(\delta, R)$-variable-length achievable rates for $\vect{X}$ is denoted by
\begin{align}
T_{\rm v} (\delta, R| {\vect{X}}, \vect{W}):= \inf\{ L : ~ L ~\mbox{is $(\delta, R)$-{variable-length achievable}~for}~\vect{X} \}. \label{eq:dR-variable_length_opt_rate}
\end{align}
When $\vect{W}$ is the identity mapping, $T_{\rm v} (\delta, R| {\vect{X}}, \vect{W})$ is simply denoted by $T_{\rm v} (\delta, R| {\vect{X}})$ (\emph{source resolvability}).
\QED
\end{e_defin}

\medskip
{Next, we {may} consider the variable-length resolvability based on the divergence.}

\begin{e_defin}[$(\delta, R)$-{Variable-Length} Channel Resolvability: Divergence]
A second-order resolution rate {$L \in (-\infty, + \infty)$} is said to be \emph{$(\delta, R)$-{variable-length achievable}} for $\vect{X}$ {(under the divergence)} where {$\delta \ge 0$} if there exists a variable-length uniform random number $U^{(L_n)}$ and
a deterministic mapping $\varphi_n : \SrAlphabet^* \rightarrow \CdAlphabet^n$ satisfying 
\begin{align}
  \limsup_{n \rightarrow \infty} \frac{1}{\sqrt{n}} \left( \E [L_n] - nR \right) &\le L, \label{eq:dR-variable_length_rate_cond2}  \\ 
  \limsup_{n \rightarrow \infty} D(\tilde{Y}^n||Y^n) &\le \delta, \label{eq:dR-variable_length_divergence_cond}   
\end{align}
where $\tilde{Y}^n $ denotes the output random variable via $W^n$ due to the input $ \tilde{X}^n = \varphi_n(U^{(L_n)})$.
The infimum of all $(\delta, R)$-variable-length achievable rates for $\vect{X}$ is denoted as
\begin{align}
T_{\rm v}^D (\delta, R| {\vect{X}}, \vect{W}):= \inf\{ L : ~ L ~\mbox{is $(\delta, R)$-{variable-length achievable}~for}~\vect{X} \}. \label{eq:dR-variable_length_opt_rate_divergence}
\end{align}
When $\vect{W}$ is the identity mapping, $T_{\rm v}^D (\delta, R| {\vect{X}}, \vect{W})$ is simply denoted by $T_{\rm v}^D (\delta, R| {\vect{X}})$  (\emph{source resolvability}).
\QED
\end{e_defin}

\begin{e_rema} \label{rema:interesting_second_order}
It is easily verified that 
\begin{align}
T_{\rm v} (\delta, R|{\vect{X}},\vect{W})  = \left\{
\begin{array}{ll}
+  \infty & \mbox{for}~ R < S_{\rm v}(\delta|{\vect{X}}, \vect{W})  \\
-  \infty & \mbox{for}~ R > S_{\rm v}(\delta|{\vect{X}}, \vect{W}).
\end{array}
\right.
\end{align}
Hence, only the case $R = S_{\rm v}(\delta|{\vect{X}}, \vect{W})$ is of our interest.
{The same remark also applies to $T_{\rm v}^D (\delta, R| {\vect{X}}, \vect{W})$}.
\QED
\end{e_rema}

\subsection{General Formulas}

We establish the general formulas for the second-order resolvability. {The proofs of the following theorems are given below subsequently to Remark \ref{rema:2nd-source-coding}}.
\begin{e_theo}[With Variational Distance]\label{theo:dR-VL_resolvability}
For any input process ${\vect{X}}$ and general channel $\vect
{W}$, 
\begin{align}
T_{\rm v} (\delta, R|{\vect{X}}, \vect{W}) = \lim_{\gamma \downarrow 0} \limsup_{n \rightarrow \infty} \frac{1}{\sqrt{n}} \left( H_{[\delta+\gamma], W^n} (\TargetSr) -n R \right)~~~( \delta \in [0, 1),  R \ge 0). \label{eq:dR-VL_ch_resolvability_formula}
\end{align}
In particular,
\begin{align}
T_{\rm v} (\delta, R|{\vect{X}}) = \lim_{\gamma \downarrow 0} \limsup_{n \rightarrow \infty} \frac{1}{\sqrt{n}} \left( H_{[\delta+\gamma]} (\TargetSr) -n R \right)~~~( \delta \in [0, 1),  R \ge 0). \label{eq:dR-VL_resolvability_formula}
\end{align}
\QED
\end{e_theo}

\begin{e_theo}[With Divergence]\label{theo:dR-mean_resolvability_div}
For any input process ${\vect{X}}$ and general channel $\vect
{W}$, 
\begin{align}
T_{\rm v}^D (\delta, R|{\vect{X}}, \vect{W}) = \lim_{\gamma \downarrow 0} \limsup_{n \rightarrow \infty} \frac{1}{\sqrt{n}} \left( H_{[\delta+\gamma], W^n}^D (\TargetSr) -n R \right)~~~( {\delta \ge 0},  R \ge 0). \label{eq:dR-VL_ch_resolvability_formula_div}
\end{align}
In particular,
\begin{align}
T_{\rm v}^D (\delta, R|{\vect{X}}) = \lim_{\gamma \downarrow 0} \limsup_{n \rightarrow \infty} \frac{1}{\sqrt{n}} \left( H_{[\delta+\gamma]}^D (\TargetSr) -n R \right)~~~( {\delta \ge 0},  R \ge 0). \label{eq:dR-VL_resolvability_formula_div}
\end{align}
\QED
\end{e_theo}

\begin{e_rema} \label{rema:2nd-mean-res}
As we have discussed in Section \ref{sec:channel_resolvability}, we may also consider to use a general source $\tilde{\vect{X}}$ as the input random variable to the channel $\vect{W}$, and we can define $L$ to be a $(\delta, R)$-\emph{mean} achievable rate for $\vect{X}$ by replacing \eqref{eq:dR-variable_length_rate_cond} and \eqref{eq:dR-variable_length_rate_cond2}  with
\begin{align}
  \limsup_{n \rightarrow \infty} \frac{1}{\sqrt{n}} \left( H(\tilde{X}^n) - nR \right) &\le L. \label{eq:dR-mean_rate_cond} 
\end{align}
Let $\overline{T}_{\rm v} (\delta, R|{\vect{X}}, \vect{W}) $ and $\overline{T}_{\rm v}^D (\delta, R|{\vect{X}}, \vect{W}) $ denote the infimum of all $(\delta, R)$-mean achievable rates for $\vect{X}$ under the variational distance and the divergence, respectively.
Then, it is not difficult to verify that
\begin{align}
\overline{T}_{\rm v} (\delta, R|{\vect{X}}, \vect{W}) &= T_{\rm v} (\delta, R|{\vect{X}}, \vect{W}) ~~~(\delta \in [0,1)), \\
\overline{T}_{\rm v}^D (\delta, R|{\vect{X}}, \vect{W}) &= T_{\rm v}^D (\delta, R|{\vect{X}}, \vect{W}) ~~~(\delta \ge 0).
\end{align}
Thus, there is no loss in the $(\delta, R)$-achievable resolution rate even if the channel input $\tilde{\vect{X}}$ is restricted to be generated by the variable-length uniform random number $U^{(L_n)}$.
\QED
\end{e_rema}

{
\begin{e_rema} \label{rema:2nd-source-coding}
As in the first-order case, when the channel $\vect{W}$ is the sequence of the identity mappings, $T_{\rm v} (\delta, R|{\vect{X}})$ coincides with the minimum second-order length rate of variable-length source codes.
More precisely, we denote by $R_{\rm v}^* (\delta, R | \vect{X})$ the \emph{minimum second-order length rate} of a sequence of variable-length source codes with  first-order average length rate $R$ and the average error probability {asymptotically not exceeding $\delta$}.
Yagi and Nomura  \cite{Yagi-Nomura2016} have shown that
\begin{align}
R_{\rm v}^* (\delta, R | \vect{X}) = \lim_{\gamma \downarrow 0} \limsup_{n \rightarrow \infty} \frac{1}{\sqrt{n}} \left( G_{[\delta+\gamma]} (\TargetSr) -n R \right)~~~( \delta \in [0, 1),  R \ge 0). \label{eq:dR-VL_coding_formula}
\end{align}  
Modifying the proof of Proposition \ref{prop:entropy_relation} (cf.\ Appendix \ref{appendix:entropy_relation}), we can show that the r.h.s.\ of \eqref{eq:dR-VL_resolvability_formula} coincides with the one  of \eqref{eq:dR-VL_coding_formula}, and therefore, it generally holds that
\begin{align}
T_{\rm v} (\delta, R | \vect{X}) = R_{\rm v}^* (\delta, R | \vect{X})~~~( \delta \in [0, 1),  R \ge 0). \label{eq:2nd-order-relation}
\end{align}
As a special case, suppose that $\vect{X}$ is a stationary and memoryless source {$X$} with the finite third absolute moment of $\log \frac{1}{P_X(X)}$.
{In this case}, Kostina et al.\ \cite{KPV2015} {have} recently given a single-letter characterization for $R_{\rm v}^* (\delta, R | \vect{X})$ with $R = H_{[\delta]}(\vect{X}) = (1 - \delta)H(X)$ as
\begin{align}
R_{\rm v}^* (\delta, R | \vect{X}) = - \sqrt{\frac{V(X)}{2 \pi}} e^{-\frac{(Q^{-1}(\delta))^2}{2}},
\end{align}
where $V(X)$ denotes the variance of $\log \frac{1}{P_X(X)}$ (varentropy) and $Q^{-1}$ is the inverse of the complementary cumulative distribution function of the standard Gaussian distribution.
In view of the general relation \eqref{eq:2nd-order-relation}, we can also obtain the single-letter characterization for $T_{\rm v} (\delta, R | \vect{X})$:
\begin{align}
T_{\rm v} (\delta, R | \vect{X}) = - \sqrt{\frac{V(X)}{2 \pi}} e^{-\frac{(Q^{-1}(\delta))^2}{2}}.
\end{align}
It has not yet been clear if we can also have a single-letter formula for $T_{\rm v} (\delta, R|{\vect{X}}, \vect{W})$ when the channel $\vect{W}$ is {memoryless but} not necessarily the identity mapping. 
\QED
\end{e_rema}
}

\medskip
\noindent
\emph{Proof of Theorems \ref{theo:dR-VL_resolvability} and \ref{theo:dR-mean_resolvability_div}}

\emph{1) Converse Part:}~~~We will show the converse part of Theorem \ref{theo:dR-VL_resolvability}. {The converse part of Theorem \ref{theo:dR-mean_resolvability_div} can be proved in {an analogous way}.} 

Let $L$ be $(\delta, R)$-{variable-length achievable} for $\vect{X}$ {under the variational distance}. 
Then, there exists $U^{(L_n)}$ and $\varphi_n$ satisfying \eqref{eq:dR-variable_length_rate_cond}  and 
\begin{align}
 &\limsup_{n \rightarrow \infty} \delta_n \le \delta, \label{eq:dR-variable_length_variational_dist_cond2}   
\end{align}
where we define $\delta_n = d(P_{Y^n}, P_{\tilde{Y}^n})$.
Equation \eqref{eq:dR-variable_length_variational_dist_cond2} implies that for any given $\gamma >0$, $\delta_n \le \delta + \gamma$ for all $n \ge n_0$ with some $n_0 > 0$, and therefore
\begin{align}
H_{[\delta + \gamma], W^n}(\TargetSr) \le H_{[\delta_n], W^n}(\TargetSr)~~~(\forall n \ge n_0). \label{eq:conv_ineq1b}
\end{align}
Since $P_{\ApproxSr} \in B_{\delta_n}(\TargetSr, W^n)$, we have
\begin{align}
H_{[\delta_n], W^n}(\TargetSr) \le H(\ApproxSr). \label{eq:conv_ineq2b}
\end{align}
On the other hand, it follows from \eqref{eq:entropy-average_length} that
\begin{align}
 H(\ApproxSr) & \le H(U^{(L_n)}) = \E [L_n] +  H(L_n), \label{eq:conv_ineq3b}
\end{align}
where the inequality is due to the fact that $\varphi_n$ is a deterministic mapping and $\ApproxSr = \varphi_n (U^{(L_n)})$.

Combining  \eqref{eq:conv_ineq1b}--\eqref{eq:conv_ineq3b} yields
\begin{align}
\limsup_{n \rightarrow \infty} \frac{1}{\sqrt{n}} \left( H_{[\delta + \gamma], W^n}(\TargetSr) -nR\right) & \le \limsup_{n \rightarrow \infty} \frac{1}{\sqrt{n}} \left( H(\ApproxSr)  -nR\right)  \nonumber \\
&\le \limsup_{n \rightarrow \infty}\frac{1}{\sqrt{n}} \left( \E [L_n]  -nR \right) +  \limsup_{n \rightarrow \infty} \frac{1}{\sqrt{n}} H(L_n) \le L, 
\end{align}
where we {have used} {\eqref{eq:length_entropy_UB}} and \eqref{eq:dR-variable_length_rate_cond} for the last inequality.
Since $\gamma > 0$ is arbitrary, we have 
\begin{align}
\lim_{\gamma \downarrow 0} \limsup_{n \rightarrow \infty} \frac{1}{\sqrt{n}} \left( H_{[\delta + \gamma], W^n}(\TargetSr) -nR\right) \le L.
\end{align}

\medskip

\emph{2) Direct Part:}~~~We will show the direct part (achievability) of Theorem \ref{theo:dR-VL_resolvability} by modifying the argument of Theorems \ref{theo:d-VL_resolvability} and \ref{theo:d-VL_ch_resolvability}. 
{The direct part of Theorem \ref{theo:dR-mean_resolvability_div} can be proved in a similar manner by {modifying} the direct part of Theorem \ref{theo:d-VL_resolvability_div} instead of Theorem \ref{theo:d-VL_resolvability}.} 

\medskip
\quad Letting
\begin{align}
L = \lim_{\gamma \downarrow 0} \limsup_{n \rightarrow \infty} \frac{1}{\sqrt{n}} \left( H_{[\delta+ \gamma], W^n}(\TargetSr) -nR \right) + {2 \gamma},
\end{align}
where $\gamma >0$ is an arbitrary constant, we shall show that $L$ is $(\delta, R)$-{variable-length achievable} for $\vect{X}$ under the variational distance.

We use the same achievability scheme as in the proof of Theorem \ref{theo:d-VL_resolvability} in slightly different parameter settings.
For $\gamma > 0$, we choose $c_n > 0$ {so} that
\begin{align}
\Pr \{ V^n \not\in T_n \} \le \gamma \label{eq:dominant_set2}
\end{align}
where  $P_{V^n} \in B_{\delta + \gamma}(\TargetSr, W^n)$ {with $H_{[\delta + \gamma], W^n} (X^n) + \gamma \ge H(V^n)$} and
\begin{align}
T_n := \left\{ \TargetSeq \in \CdAlphabet^n : \frac{1}{n} \log \frac{1}{P_{V^n}(\TargetSeq)} \le c_n \right\}.
\end{align}
We here define
\begin{align}
\ell (\TargetSeq) := \left\{
\begin{array}{ll}
 \lceil \log \frac{1}{P_{V^n}(\TargetSeq)} + \sqrt{n} \gamma \rceil & \mbox{for}~ \TargetSeq \in T_n \\
 0 & \mbox{otherwise} 
\end{array} 
\right. 
\end{align}
and $\beta_n =  \lceil  (n c_n+ \sqrt{n} \gamma) \rceil$.
Arguing similarly to the proof  of Theorems \ref{theo:d-VL_resolvability} and \ref{theo:d-VL_ch_resolvability}, we can show that there exists $\varphi_n:\SrAlphabet^* \rightarrow \CdAlphabet^n$ and $U^{(L_n)}$ such that
\begin{align}
 d(P_{Y^n}, P_{\tilde{Y}^n}) \le \delta + 2 \gamma + \frac{1}{2} K^{-\sqrt{n}\gamma} \label{eq:distance_UB2}
\end{align}
and
\begin{align}
\E[L_n]  &\le \left( 1 + \frac{1}{K^{\sqrt{n}\gamma}} \right) \left( H_{[\delta+ \gamma], W^n}(\TargetSr) + 2\sqrt{n}\gamma + 1 \right). \label{eq:average_length2}
\end{align}
Therefore, we obtain
\begin{align}
\lim_{n\rightarrow \infty} d(P_{Y^n}, P_{\tilde{Y}^n})  \le \delta + 2 \gamma \label{eq:distance_UB3}
\end{align}
and
\begin{align}
 \limsup_{n \rightarrow \infty} \frac{1}{\sqrt{n}} \left( \E[L_n] - nR \right)  &\le \limsup_{n \rightarrow \infty} \frac{1}{\sqrt{n}}  \left( H_{[\delta+ \gamma], W^n}(\TargetSr) -nR \right) + 2 \gamma \nonumber \\
  &\le  \lim_{\gamma \downarrow 0} \limsup_{n \rightarrow \infty} \frac{1}{\sqrt{n}}  \left( H_{[\delta+ \gamma], W^n}(\TargetSr) -nR \right) + 2 \gamma = L.  \label{eq:average_length3} 
\end{align}
Since $\gamma > 0$ is arbitrary, $L$ is $(\delta, R)$-{variable-length achievable} for $\vect{X}$.

\medskip\section{Conclusions} \label{sec:conclusion}

\begin{table}[t]
  \caption{{Summary of {First-Order} Resolvability and Information Quantities}}
  \label{table:summary}
  \centering
  \begingroup
\renewcommand{\arraystretch}{1.5}
  \begin{tabular}{c|c|c|c}
    \hline
    Approximation Measure  & Resolvability  &  Characterization  &  Theorem \# \\
    \hline \hline
    \multicolumn{4}{c}{\emph{Fixed-Length Resolvability}} \\
    \hline
    \multirow{2}{*}{Variational Distance}  & $S_{\rm f} ({\vect{X}})$  & $\overline{H}({\vect{X}})$ & Theorem \ref{theo:fixed_length_resolvability} (\cite{Han-Verdu93}) \\
      & $S_{\rm f} (\delta| {\vect{X}})$   & $\overline{H}_\delta ({\vect{X}})$ &  Theorem \ref{theo:d-fixed_length_resolvability} (\cite{Steinberg-Verdu96}) \\
  \hline 
  \multicolumn{4}{c}{\emph{Variable-Length Resolvability}} \\
    \hline
     \multirow{3}{*}{\vspace*{-3mm} Variational Distance}  & $ S_{\rm v} ({\vect{X}}) $   & $\displaystyle \lim_{\gamma \downarrow 0} G_{[\gamma]} ({\vect{X}}) =\lim_{\gamma \downarrow 0} H_{[\gamma]} ({\vect{X}})  $ &   Theorem \ref{theo:mean_resolvability}\\
       & $S_{\rm v} (\delta|{\vect{X}}) $   & $\displaystyle \lim_{\gamma \downarrow 0} G_{\delta+ \gamma]} ({\vect{X}}) =\lim_{\gamma \downarrow 0} H_{[\delta+ \gamma]} ({\vect{X}})  $ &   Theorem \ref{theo:d-VL_resolvability}\\
      & $S_{\rm v} (\delta | \vect{X}, \vect{W})  $   & $\displaystyle \lim_{\gamma \downarrow 0} H_{[\delta+\gamma], \vect{W}}(\vect{X}) $ &   Theorem \ref{theo:d-VL_ch_resolvability}\\
    \hline
     \multirow{2}{*}{\vspace*{-3mm}  Divergence}  & $S_{\rm v}^D (\delta|{\vect{X}}) $   & $\displaystyle \lim_{\gamma \downarrow 0} H_{[\delta+ \gamma]}^D ({\vect{X}})  $ &   Theorem \ref{theo:d-VL_resolvability_div}\\
    & $S_{\rm v}^D (\delta | \vect{X}, \vect{W})  $   & $\displaystyle \lim_{\gamma \downarrow 0} H_{[\delta+\gamma], \vect{W}}^D (\vect{X}) $ &   Theorem \ref{theo:d-VL_ch_resolvability_div} \\
    \hline
  \end{tabular}
  \endgroup
\end{table}

We have investigated the problem of \emph{variable-length} source/channel resolvability, where a given target probability distribution is approximated by transforming \emph{variable-length uniform random numbers}. 
{Table \ref{table:summary} summarizes the various {
first-order} resolvability and its characterization by an information quantity. In this table, the theorem numbers which contain the corresponding characterization are also indicated.} 

{In this paper,} we have first analyzed the fundamental limit on the variable-length {$\delta$}-source resolvability with the variational distance in Theorems \ref{theo:mean_resolvability} and \ref{theo:d-VL_resolvability}.
The variable-length {$\delta$}-source resolvability is essentially characterized in terms of the smooth R\'enyi entropy of order one. 
In the proof of the direct part, we have developed a simple method of information spectrum slicing, in which sliced information densities quantized to {the same} integer are approximated by {a fixed-length} uniform random number of the same length. 
Next, we have extended the analysis to the $\delta$-source resolvability under the unnormalized divergence in Theorem \ref{theo:d-VL_resolvability_div}.
The smoothed entropy with the divergence again plays an important role in characterizing the $\delta$-source resolvability. 

Then, we have addressed the problem of {$\delta$}-channel resolvability.
It has been revealed in Theorems \ref{theo:d-VL_ch_resolvability} and \ref{theo:d-VL_ch_resolvability_div} that using an arbitrary general source as a coin distribution (mean-resolvability problem) cannot go beyond the fundamental limit of the variable-length resolvability, in which only variable-length uniform random numbers are allowed to be a coin distribution.
As in the case of source resolvability, we have discussed the $\delta$-channel resolvability under the variational distance and the unnormalized divergence.
The second-order channel resolvability has been characterized  in Theorems \ref{theo:dR-VL_resolvability} and \ref{theo:dR-mean_resolvability_div} as well as the first-order case.
When the variational distance is used as {an approximation measure}, it turned out that the $\delta$-\emph{source} resolvability is equal to the minimum achievable rate by variable-length source codes with the error probability less than or equal to $\delta$. 
This is a parallel relationship between \emph{fixed-length source resolvability} and the minimum achievable rate by \emph{fixed-length source codes} \cite{Han-Verdu93,Steinberg-Verdu96}.
It is of interest to investigate if there is a coding problem for which the $\delta$-\emph{channel} resolvability 
 is closely related.

When $\delta =0$, the asymptotically exact approximation is required. In the case where the channel $\vect{W}$ is the sequence of identity mappings, it turned out that the source resolvability under the variational distance and the unnormalized divergence coincides and is given by $\lim_{\gamma \downarrow 0} H_{[\gamma]}(\vect{X})$, where $\vect{X}$ is the target general source. This result is analogous to the dual problem of \emph{variable-length intrinsic randomness} \cite{Han2000,Vembu-Verdu95}, in which the maximum achievable rates of {variable-length} uniform random numbers extracted from a given source $\vect{X}$ are the same under the two kinds of {approximation} measures.
It should be emphasized that in the case of variable-length intrinsic randomness, the use of \emph{normalized} divergence as {an approximation measure} results in the same general formula as with the variational distance and the unnormalized divergence, which does not necessarily holds in the case of mean/variable-length resolvability (cf.\ Remark \ref{rema:normalized_divergence}).    
It is also noteworthy that whereas only the case of $\delta =0$ has been completely solved for the variable-length intrinsic randomness, we have also dealt with the case $\delta > 0$ for the variable-length source/channel resolvability.

When $\vect{X}$ is a stationary and memoryless source, the established formulas reduce to a single letter characterization for the first- and second-order source resolvability under the variational distance. 
In the case where the divergence is the {approximation} measure and/or the channel $\vect{W}$ is a non-identity mapping, however, it has not yet been clear if we can derive a single-letter characterization for the $\delta$-source/channel resolvability. This question remains open to be studied.

\appendices

\medskip
\medskip\section{Proof of Lemma \ref{lem:entropy_closeness}} \label{appendix:entropy_closeness_lemma}

Letting $\delta_n = {2}\, {d(P_{\TargetSr}, P_{\ApproxSr})}$, we define 
\begin{align}
T_n = \left\{ \TargetSeq \in \CdAlphabet^n: \left| 1 - \frac{P_{\ApproxSr}(\TargetSeq)}{P_{\TargetSr}(\TargetSeq)}\right| \le \sqrt{\delta_n}\right\}.  
\end{align}
Then,  by using Markov's inequality we obtain
\begin{align}
\Pr\{ \TargetSr \in T_n \} &\ge 1 - \sqrt{\delta_n}  \label{eq:prob_Y_in_setT}
\end{align}
as shown by Han \cite[Proof of Theorem 2.1.3]{Han2003}.
It should be noticed that for any $\TargetSeq \in T_n$ it holds that
\begin{align}
1 - \sqrt{\delta_n} \le \frac{P_{\ApproxSr}(\TargetSeq)}{P_{\TargetSr}(\TargetSeq)} \le  1 + \sqrt{\delta_n},  \label{eq:LLR_UB1}
\end{align}
and therefore
\begin{align}
\frac{1}{n} \log \frac{1}{P_{\TargetSr}(\TargetSeq)} \le \frac{1}{n} \log \frac{1}{P_{\ApproxSr}(\TargetSeq)} + \frac{1}{n} \log (1 + \sqrt{\delta_n}). \label{eq:LLR_UB2}
\end{align}

Fix $\gamma \in (0, 1)$ arbitrarily. 
We choose a subset $A_n \subseteq \CdAlphabet^n$ satisfying
\begin{align}
G_{[\delta+\gamma]}(\ApproxSr) + \gamma &\ge \sum_{\TargetSeq \in A_n} P_{\ApproxSr}(\TargetSeq) \log \frac{1}{P_{\ApproxSr}(\TargetSeq)}, \label{eq:dominant_set_An_entropy} \\
\Pr\{\ApproxSr \in A_n \} &\ge 1 - \delta - \gamma. \label{eq:dominant_set_An_prob} 
\end{align}
By the definition of variational distance, {it is well-known} that
\begin{align}
\delta_n &= {2} \sup_{S_n \subseteq \CdAlphabet^n} |\Pr\{ \TargetSr \in S_n\} - \Pr\{ \ApproxSr \in S_n\}| \nonumber \\
&\ge  |\Pr\{ \TargetSr \in A_n\} - \Pr\{ \ApproxSr \in A_n\}| , \label{eq:variational_dist_property1}
\end{align}
and therefore \eqref{eq:dominant_set_An_prob} indicates that
\begin{align}
\Pr\{\TargetSr \in A_n \} \ge  1 - \delta - \gamma - \delta_n.
\end{align}
Defining $B_n = A_n \cap T_n$, we also have 
\begin{align}
\Pr\{\TargetSr \in B_n \} \ge \Pr\{ \TargetSr \in A_n \} - \Pr \{ \TargetSr \not\in T_n \}  \ge 1 - \delta - \gamma - \delta_n - \sqrt{\delta_n}, \label{eq:prob_Y_in_B}
\end{align}
where the last inequality follows from \eqref{eq:prob_Y_in_setT}.
On the other hand, it follows from \eqref{eq:LLR_UB1} and \eqref{eq:LLR_UB2} that
\begin{align}
\sum_{\TargetSeq \in B_n} P_{\TargetSr}(\TargetSeq) \cdot \frac{1}{n} \log \frac{1}{P_{\TargetSr}(\TargetSeq)} \le \frac{1}{1 - \sqrt{\delta_n}}\sum_{\TargetSeq \in B_n} P_{\ApproxSr}(\TargetSeq) \cdot \left( \frac{1}{n} \log \frac{1}{P_{\ApproxSr}(\TargetSeq)} + \frac{1}{n} \log (1 + \sqrt{\delta_n}) \right). \label{eq:partial_entropy_relation1}
\end{align}
Then, in view of \eqref{eq:prob_Y_in_B} and the definition of $G_{[\delta]}({\vect{X}})$, it is easily checked that we have
\begin{align}
G_{[\delta+ 2 \gamma]} ({\vect{X}}) &\le \limsup_{n \rightarrow \infty} \frac{1}{n} \sum_{\TargetSeq \in B_n} P_{\TargetSr}(\TargetSeq) \log \frac{1}{P_{\TargetSr}(\TargetSeq)} \nonumber \\
&\le  \limsup_{n \rightarrow \infty} \frac{1}{n} \sum_{\TargetSeq \in B_n} P_{\ApproxSr}(\TargetSeq) \log \frac{1}{P_{\ApproxSr}(\TargetSeq)} \nonumber \\
&\le  \limsup_{n \rightarrow \infty} \frac{1}{n} \sum_{\TargetSeq \in A_n} P_{\ApproxSr}(\TargetSeq) \log \frac{1}{P_{\ApproxSr}(\TargetSeq)} \le G_{[\delta+ \gamma]} (\tilde{{\vect{X}}}),
\end{align}
where the last inequality is due to \eqref{eq:dominant_set_An_entropy}.
Since $\gamma \in (0,1)$ is arbitrary, taking $\gamma \downarrow 0$ for both sides yields
\begin{align}
\lim_{\gamma \downarrow 0} G_{[\delta+ \gamma]} ({\vect{X}}) \le \lim_{\gamma \downarrow 0} G_{[\delta+ \gamma]} (\tilde{{\vect{X}}}). \label{eq:entropy_closeness1}
\end{align}
By the symmetry of the argument, \eqref{eq:entropy_closeness1} indicates \eqref{eq:entropy_closeness}.

\medskip\section{Proof of Equation \eqref{eq:smooth_entropy_equivalence}} \label{append:smooth_entropy_equivalence}

\medskip

\noindent
(i)~~We first show $\lim_{\alpha \uparrow 1} H_{[\delta]}^\alpha(\TargetSr) \le H_{[\delta]}(\TargetSr)$.

Fix $\gamma > 0$ arbitrarily. We choose $P_{V^n} \in B_{\delta}(\TargetSr)$ satisfying
\begin{align}
H_{[\delta]}(\TargetSr) + \gamma \ge H(V^n). \label{eq:ineq151}
\end{align}
It is well known that {the} R\'enyi entropy of order $\alpha$ defined as
\begin{align}
H^\alpha(V^n)=\frac{1}{1 - \alpha} \log \sum_{\TargetSeq \in \CdAlphabet^n} P_{V^n}(\TargetSeq)^\alpha~~~(\forall \alpha \in (0,1) \cup (1, +\infty)) \nonumber
\end{align} satisfies
\begin{align}
H(V^n) = \lim_{\alpha \rightarrow 1} H^\alpha(V^n). \label{eq:ineq152}
\end{align} 
By definition, we have
\begin{align}
H^\alpha(V^n) \ge H_{[\delta]}^\alpha(\TargetSr)~~~(\forall \alpha  \in (0,1) \cup (1, +\infty)),  \label{eq:ineq153}
\end{align}
leading to 
\begin{align}
H(V^n)= \lim_{\alpha \uparrow 1} H^\alpha(V^n) \ge  \lim_{\alpha \uparrow 1} H_{[\delta]}^\alpha(\TargetSr).  \label{eq:ineq153b}
\end{align}
Combining {\eqref{eq:ineq151} and \eqref{eq:ineq153b} yields}
\begin{align}
H_{[\delta]}(\TargetSr) + \gamma \ge H(V^n) \ge \lim_{\alpha \uparrow 1} H_{[\delta]}^\alpha(\TargetSr). 
\end{align}
Since $\gamma > 0$ is an arbitrary constant, we obtain $\lim_{\alpha \uparrow 1} H_{[\delta]}^\alpha(\TargetSr) \le H_{[\delta]}(\TargetSr)$.

\medskip
\noindent
(ii)~~Next, we shall show $\lim_{\alpha \uparrow 1} H_{[\delta]}^\alpha(\TargetSr) \ge H_{[\delta]}(\TargetSr)$.

Fix $\gamma > 0$ arbitrarily. 
We choose some $\alpha_0 \in (0,1)$ satisfying 
\begin{align}
\lim_{\alpha \uparrow 1} H_{[\delta]}^\alpha(\TargetSr) + \gamma \ge H_{[\delta]}^{\alpha_0} (\TargetSr). \label{eq:ineq155}
\end{align}
For this $\alpha_0$ we choose $P_{V^n} \in B_\delta(\TargetSr)$ satisfying 
\begin{align}
H_{[\delta]}^{\alpha_0} (\TargetSr) +  \gamma \ge H^{\alpha_0} (V^n). \label{eq:ineq156}
\end{align}
Since $H^\alpha(V^n)$ is a nonincreasing function {of} $\alpha$, we have
\begin{align}
H^{\alpha_0} (V^n) \ge H(V^n),  \label{eq:ineq157}
\end{align}
and it follows from \eqref{eq:ineq155}--\eqref{eq:ineq157} that 
\begin{align}
\lim_{\alpha \uparrow 1} H_{[\delta]}^\alpha(\TargetSr) + 2 \gamma \ge H^{\alpha_0} (V^n) \ge H(V^n). \label{eq:ineq158}
\end{align}
Since $H(V^n) \ge H_{[\delta]}(\TargetSr)$ due to $P_{V^n} \in B_\delta(\TargetSr)$ and $\gamma>0$ is arbitrarily fixed, we obtain the desired inequality.

\medskip\section{Proof of Equation \eqref{eq:d-mean_resolvability_formula_alt}} \label{append:formula_alt}

\medskip
To prove the alternative formula \eqref{eq:d-mean_resolvability_formula_alt} for the $\mathrm{v}(\delta)$-resolvability $S_{\rm v}(\delta|\vect{X})$, we shall show 
\begin{align}
\lim_{\gamma \downarrow 0} H_{[\delta+ \gamma]}(\vect{X}) = \inf_{\vect{V} \in B_\delta(\vect{X})}H(\vect{V})~~~(\delta \in [0,1)).
\end{align}

\noindent
(i)~~We first show $\displaystyle \lim_{\gamma \downarrow 0} H_{[\delta+ \gamma]}(\vect{X}) \le \inf_{\vect{V} \in B_\delta(\vect{X})}H(\vect{V})$.

Fix $\gamma > 0$ arbitrarily. We choose $\tilde{\vect{V}} = \{ \tilde{V}^n\}_{n=1}^\infty \in B_\delta (\vect{X})$ satisfying
\begin{align}
H(\tilde{\vect{V}}) \le \inf_{\vect{V} \in B_\delta(\vect{X})}H(\vect{V}) + \gamma. \label{eq:ineq31}
\end{align}
For $\tilde{\vect{V}} \in B_\delta (\vect{X})$, we have $d(X^n , \tilde{V}^n) \le \delta + \gamma$ for all $n \ge n_0$ with some $n_0 > 0$, yielding
\begin{align}
H_{[\delta + \gamma]}(X^n) \le H(\tilde{V}^n) ~~~(\forall n \ge n_0). \label{eq:ineq32}
\end{align}
Thus, it follows from \eqref{eq:ineq31} and \eqref{eq:ineq32} that 
\begin{align}
 H_{[\delta + \gamma]}(\vect{X}) \le \inf_{\vect{V} \in B_\delta(\vect{X})}H(\vect{V}) + \gamma.
\end{align}
Since $\gamma > 0$ is an arbitrary constant, letting $\gamma \downarrow 0$ on both sides yields the desired inequality.

\medskip
\noindent
(ii)~~Next, we shall show $\displaystyle \lim_{\gamma \downarrow 0} H_{[\delta+ \gamma]}(\vect{X}) \ge \inf_{\vect{V} \in B_\delta(\vect{X})}H(\vect{V})$.

Fix $\lambda > 0$ arbitrarily. We choose an arbitrary decreasing sequence of positive numbers $\{ \gamma_i \}_{i=1}^\infty$ satisfying $\gamma_1 > \gamma_2> \cdots \rightarrow 0$.
Then, we have 
\begin{align}
\lim_{\gamma \downarrow 0} H_{[\delta+ \gamma]}(\vect{X}) = \lim_{i \rightarrow \infty} H_{[\delta+ \gamma_i]}(\vect{X}). \label{eq:ineq33a} 
\end{align}
Also, by the definition of the limit superior, for each $i=1, 2, \cdots$ we have
\begin{align}
\frac{1}{n} H_{[\delta + \gamma_i]} (X^n) \le H_{[\delta+ \gamma_i]} (\vect{X}) + \lambda~~~(\forall n \ge n_i) \label{eq:ineq33b} 
\end{align}
{with some $0 < n_1 < n_2 < \cdots$}.
Now, for each $n=1, 2, \cdots$, we denote by $i_n$ the index $i$ satisfying
\begin{align}
n_i \le n < n_{i+1}.
\end{align}
Then from \eqref{eq:ineq33b}, we obtain
\begin{align}
\frac{1}{n} H_{[\delta + \gamma_{i_n}]} (X^n) \le H_{[\delta+ \gamma_{i_n}]} (\vect{X}) + \lambda~~~(\forall n \ge n_1). \label{eq:ineq34} 
\end{align}

On the other hand, by the definition of $H_{[\delta + \gamma_{i_n}]} (X^n)$, for each $n=1, 2, \cdots$, we can choose some $V_{i_n}^n \in B_{\delta + \gamma_{i_n}}(X^n)$ satisfying
\begin{align}
\frac{1}{n} H (V_{i_n}^n) \le \frac{1}{n} H_{[\delta + \gamma_{i_n}]} (X^n) + \lambda. \label{eq:ineq35} 
\end{align}
We {now} construct the general source $\tilde{\vect{V}} = \{ V_{i_n}^n\}_{n=1}^\infty$ from each $V_{i_n}^n$ for $n=1, 2, \cdots$.
Since $V_{i_n}^n \in B_{\delta + \gamma_{i_n}}(X^n)$ for all  $n \ge n_1$ indicates that
\begin{align}
\limsup_{n \rightarrow \infty} d(X^n, V_{i_n}^n) \le \delta + \lim_{n\rightarrow \infty} \gamma_{i_n} = \delta,
\end{align}
the general source satisfies $\tilde{\vect{V}} \in B_\delta(\vect{X})$.

From \eqref{eq:ineq34}  and \eqref{eq:ineq35}, we obtain
\begin{align}
\frac{1}{n} H (V_{i_n}^n)  \le H_{[\delta+ \gamma_{i_n}]} (\vect{X}) + 2\lambda~~~(\forall n \ge n_1). \label{eq:ineq36} 
\end{align}
In view of \eqref{eq:ineq33a} and the fact $\tilde{\vect{V}} \in B_\delta(\vect{X})$, taking $\displaystyle \limsup_{n \rightarrow \infty}$ on both sides yields
\begin{align}
\inf_{\vect{V} \in B_\delta(\vect{X})}H(\vect{V}) \le H (\tilde{\vect{V}})  &\le \limsup_{n \rightarrow \infty} H_{[\delta+ \gamma_{i_n}]} (\vect{X}) + 2\lambda \nonumber \\
&=  \lim_{\gamma \downarrow 0} H_{[\delta+ \gamma]} (\vect{X}) + 2\lambda.  \label{eq:ineq37} 
\end{align}
Since $\lambda > 0$ is an arbitrary constant, letting $\lambda \downarrow 0$ yields the desired inequality.

\medskip\section{Proof of Proposition \ref{prop:entropy_relation}} \label{appendix:entropy_relation}

Equation \eqref{eq:asymptotic_equivalence3} is an immediate consequence of \eqref{eq:asymptotic_equivalence2} because $\overline{H}_{\delta}(\vect{X})$ is a right-continuous function \cite{Han2003}.
The rightmost inequality in \eqref{eq:asymptotic_equivalence2} is due to \cite{Han2015}, which improves a bound established in \cite{Koga-Yamamoto2005} and \cite{Kuzuoka-Watanabe2015}.
We shall show the leftmost {equality}: $H_{[\delta]} ({\vect{X}}) =  G_{[\delta]} ({\vect{X}}) $.

\medskip
\noindent
(i) We first show $H_{[\delta]} ({\vect{X}}) \le  G_{[\delta]} ({\vect{X}})$.

For any given $\gamma > 0$ and $P_{\TargetSr}$, let $\mathcal{A}_n^* \subseteq \CdAlphabet^n$ be a subset of $\CdAlphabet^n$ which satisfies 
\begin{align}
\Pr\{\TargetSr \in \mathcal{A}_n^* \} \ge 1 - \delta \label{eq:prob_cond}
\end{align}
and 
\begin{align}
G_{[\delta]}(\TargetSr)  + \gamma \ge  \sum_{\TargetSeq \in \mathcal{A}_n^*} P_{\TargetSr}(\TargetSeq) \log \frac{1}{P_{\TargetSr}(\TargetSeq)} =: {F}(\mathcal{A}_n^*). \label{eq:func_g}
\end{align}
Choose $\TargetSeq_0 \in \CdAlphabet^n \setminus \mathcal{A}_n^*$ arbitrarily.
Setting $P_{\ApproxSr}$ such that
\begin{align}
P_{\ApproxSr}(\TargetSeq) = \left\{
\begin{array}{ll}
P_{\TargetSr}(\TargetSeq) & \mbox{for}~ \TargetSeq \in \mathcal{A}_n^* \\
\alpha_0 & \mbox{for}~ \TargetSeq = \TargetSeq_0 \\
0 & \mbox{otherwise}, 
\end{array}
 \right.
 \end{align}
 where we define $\alpha_0 = \Pr\{ \TargetSr \not\in \mathcal{A}_n^* \}$.

The variational distance between $P_{\TargetSr}$ and $P_{\ApproxSr}$ satisfies
\begin{align}
d(P_{\TargetSr}, P_{\ApproxSr}) &= \frac{1}{2} \sum_{\TargetSeq \not\in \mathcal{A}_n^*} |P_{\TargetSr}(\TargetSeq) -  P_{\ApproxSr}(\TargetSeq)| \nonumber \\
& \le \frac{1}{2} \sum_{\TargetSeq \not\in \mathcal{A}_n^*} ( P_{\TargetSr}(\TargetSeq) + P_{\ApproxSr}(\TargetSeq) ) \le \alpha_0  = 1 - \Pr\{ \TargetSr \in \mathcal{A}_n^* \} \le \delta, 
\end{align}
where the last inequality is due to \eqref{eq:prob_cond}.
Therefore, $P_{\ApproxSr} \in B_\delta (\TargetSr)$, and this implies
\begin{align}
H_{[\delta]} (\TargetSr) \le H(\ApproxSr) &= {F}(\mathcal{A}_n^*) + \alpha_0 \log \frac{1}{\alpha_0} \nonumber \\
 & \le  G_{[\delta]}  (\TargetSr) + \alpha_0 \log \frac{1}{\alpha_0} + \gamma \nonumber \\
 & \le  G_{[\delta]}  (\TargetSr) + \frac{\log e}{e} + \gamma,
 \end{align}
 where the first inequality is due to \eqref{eq:func_g} and the last inequality is due to the inequality $x \log x \ge - \frac{\log e}{e}$ for all $x >0$.
Thus, we obtain the desired inequality: $H_{[\delta]} ({\vect{X}}) \le  G_{[\delta]} ({\vect{X}}) $.

\medskip
\noindent
{(ii) Next, we shall show $H_{[\delta]} ({\vect{X}}) \ge  G_{[\delta]} ({\vect{X}})$.}

{
Assume, without loss of generality, that the elements of $\mathcal{X}^n$ are indexed as $\vect{x}_1, \vect{x}_2, \cdots \in \mathcal{X}^n$ {so that}
\begin{align}
P_{X^n}(\vect{x}_i) \ge P_{X^n}(\vect{x}_{i+1})~~~(\forall i = 1, 2, \cdots).
\end{align}
For a given $\delta \in [0,1)$, let $j^*$ denote the integer satisfying
\begin{align}
\sum_{i=1}^{j^* - 1} P_{X^n}(\vect{x}_i) < 1 - \delta,~~~~~\sum_{i=1}^{j^*} P_{X^n}(\vect{x}_i) \ge 1 - \delta.
\end{align}
Let $V_\delta^n$ be a random variable taking values in $\mathcal{X}^n$ whose probability distribution is given by
\begin{align}
P_{V_\delta^n} (\vect{x}_i) = \left\{ 
\begin{array}{ll}
P_{X^n}(\vect{x}_i) + \delta & \mathrm{for}~ i =1 \\
P_{X^n}(\vect{x}_i)  & \mathrm{for}~ i =2, 3, \cdots, j^* -1 \\
P_{X^n}(\vect{x}_i)  - \varepsilon_n &  \mathrm{for}~ i =j^*\\
 0 &  \mathrm{otherwise},
\end{array}
\right.
\end{align}
where we define $\varepsilon_n = \delta - \sum_{i \ge j^*+1} P_{X^n}(\vect{x}_i)$.
It is easily checked that {$0 \le \varepsilon_n \le P_{X^n}(\vect{x}_{j^*})$ and} the probability distribution $P_{V_\delta^n}$ \emph{majorizes}\footnote{For a sequence $\vect{u}= (u_1, u_2, \cdots, u_L)$ of length $L$, we denote by $\tilde{\vect{u}} = (\tilde{u}_1, \tilde{u}_2, \cdots, \tilde{u}_L)$ the permuted version of $\vect{u}$ satisfying $\tilde{u}_i \ge \tilde{u}_{i+1}$ for all $i = 1, 2, \cdots, L$, where ties are arbitrarily broken.
We say $\vect{u}= (u_1, u_2, \cdots, u_L)$ \emph{majorizes} $\vect{v}= (v_1, v_2, \cdots, v_L)$ if $\sum_{i=1}^j \tilde{u}_i \ge \sum_{i=1}^j \tilde{v}_i$ for all $j=1, 2, \cdots, L$.} any $P_{V^n} \in B_\delta(X^n)$ \cite{Ho-Yeung2010}.
Since the Shannon entropy is a \emph{Schur concave} function\footnote{A function $f(\vect{u})$ is said to be \emph{Schur concave} if  $f(\vect{u}) \le f(\vect{v})$ for any pair $(\vect{u}, \vect{v})$, where $\vect{v}$ is majorized by $\vect{u}$.}  \cite{MOA2011}, we immediately obtain the following lemma, which is of use to compute $H_{[\delta]}(X^n)$.
\begin{e_lem}[Ho and Yeung \cite{Ho-Yeung2010}] \label{lem:smooth_entropy_compute}
\begin{align}
H_{[\delta]}(X^n) = H(V_\delta^n) ~~~(\forall \delta \in [0,1)).
\end{align}
\QED 
\end{e_lem}}

{
By the definition of $G_{[\delta]}(X^n)$, we obtain
\begin{align}
G_{[\delta]}(X^n) &\le \sum_{i=1}^{j^*} P_{X^n}(\vect{x}_i) \log \frac{1}{P_{X^n}(\vect{x}_i)} \\
 &\le H(V_\delta^n) + P_{X^n}(\vect{x}_1) \log \frac{1}{P_{X^n}(\vect{x}_1)} + P_{X^n}(\vect{x}_{j^*}) \log \frac{1}{P_{X^n}(\vect{x}_{j^*})} \\
 &\le  H(V_\delta^n) + \frac{2 \log e}{ e},
\end{align}
where the last inequality is due to $x \log x \ge - \frac{\log e}{e}$ for all $x > 0$.
Thus, it follows from Lemma \ref{lem:smooth_entropy_compute} that
\begin{align}
G_{[\delta]}(\vect{X}) &\le \limsup_{n \rightarrow \infty} \frac{1}{n} H(V_\delta^n) = H_{[\delta]}(\vect{X}),
\end{align}
which is the desired inequality.}

\medskip\section{Proof of Lemma \ref{lem:entropy_relation_divergence}} \label{appendix:proof_entropy_relation_divergence}

We first show \eqref{eq:asymptotic_equivalence4}.
For  two general sources ${\vect{X}} = \{ \TargetSr \}_{n=1}^\infty$ and $\tilde{{\vect{X}}} = \{ \ApproxSr \}_{n=1}^\infty$, the following well-known inequality (cf.\ \cite[Problem 3.18]{Csiszar-Korner2011}) holds between {the} variational distance and {the} divergence:
\begin{align}
{\frac{2 \big(d(P_{\TargetSr}, P_{\ApproxSr}) \big)^2}{\ln K}} \le D(\ApproxSr||\TargetSr).  \label{eq:Pinsker_ineq}
\end{align}
This inequality implies that any $P_{V^n} \in B_{g(\delta)}^D (\TargetSr)$ satisfies
$P_{V^n} \in B_{\delta} (\TargetSr)$.
Thus, we have
\begin{align}
H_{[\delta]} (\TargetSr) \le H_{[g(\delta)]}^D(\TargetSr).
\end{align}

Now, we shall show {the rightmost equality of} \eqref{eq:asymptotic_equivalence5}. It obviously follows from \eqref{eq:asymptotic_equivalence4} that
\begin{align}
\lim_{\delta \downarrow 0} H_{[\delta]}({\vect{X}}) \le \lim_{\delta \downarrow 0} H_{[\delta]}^D ({\vect{X}}).  
\end{align}
To show the opposite inequality, in view of \eqref{eq:asymptotic_equivalence3} it suffices to show
\begin{align}
\lim_{\delta \downarrow 0} G_{[\delta]}({\vect{X}}) \ge \lim_{\delta \downarrow 0} H_{[\delta]}^D ({\vect{X}}). \label{eq:ineq20}
\end{align}
Fix $\delta \in (0,1)$ and $\gamma >0$ arbitrarily. We choose $\mathcal{A}_n \subseteq \CdAlphabet^n$ satisfying
\begin{align}
G_{[\delta]} (\TargetSr) + \gamma &\ge \sum_{\TargetSeq \in \mathcal{A}_n} P_{\TargetSr}(\TargetSeq) \log \frac{1}{P_{\TargetSr}(\TargetSeq)},  \label{eq:ineq21} \\
\alpha_0 &:= \Pr\{ \TargetSr \in \mathcal{A}_n \} \ge 1 - \delta. \label{eq:ineq22}
\end{align}
We arrange a new random variable $V^n$ subject to 
\begin{align}
P_{V^n}(\TargetSeq) = \left\{
\begin{array}{ll}
\frac{P_{\TargetSr}(\TargetSeq)}{\alpha_0} & \mbox{if}~ \TargetSeq \in \mathcal{A}_n \\
0 & \mbox{otherwise}.
\end{array}
\right.
\end{align}
Then, we obtain
\begin{align}
D(V^n || \TargetSr) = \sum_{\TargetSeq \in \mathcal{A}_n} P_{V^n} (\TargetSeq) \log \frac{P_{V^n}(\TargetSeq)}{P_{\TargetSr}(\TargetSeq)} = \log \frac{1}{\alpha_0} \le \log \frac{1}{1 - \delta},
\end{align}
and thus letting $h(\delta) = \log \frac{1}{1-\delta}$, it holds that $P_{V^n} \in B_{h(\delta)}^D (\TargetSr)$.
We can expand \eqref{eq:ineq21} as
\begin{align}
G_{[\delta]} (\TargetSr) + \gamma &\ge \alpha_0  \sum_{\TargetSeq \in \mathcal{A}_n} P_{V^n}(\TargetSeq) \log \frac{1}{\alpha_0 P_{V^n}(\TargetSeq)} \nonumber \\
&\ge  \alpha_0 H(V^n) \nonumber \\
&\ge (1-\delta) H_{[h(\delta)]}^D(\TargetSr),  \label{eq:ineq23}
\end{align}
where the last inequality is due to \eqref{eq:ineq22} and $P_{V^n} \in B_{h(\delta)}^D (\TargetSr)$. 
Thus, {as $\gamma> 0$ is arbitrary},
\begin{align}
G_{[\delta]} ({\vect{X}}) \ge   (1-\delta) H_{[h(\delta)]}^D({\vect{X}}). 
\end{align}
Since $\delta \in (0,1)$ is arbitrary, in view of $\lim_{\delta \downarrow 0} h(\delta) = 0$ we obtain \eqref{eq:ineq20}.
\QED

\end{document}